\documentclass[preprint, tighten, twocolumn]{aastex63}
\usepackage{apjfonts}
\usepackage{amsmath}
\usepackage{multirow}
\usepackage{textcomp}
\usepackage{appendix}
\usepackage{color}

\shorttitle{Black Hole Masses in NGC 315 and NGC 4261}
\shortauthors{Boizelle et al.}

\usepackage{natbib}
\defcitealias{boizelle17}{Paper I}
\defcitealias{boizelle19}{Paper II}

\newcommand{\hst}{HST}
\newcommand{\spitzer}{Spitzer}
\newcommand{\kms}{km s\ensuremath{^{-1}}}
\newcommand{\vlos}{\ensuremath{v_\mathrm{LOS}}}

\newcommand{\vsys}{\ensuremath{v_\mathrm{sys}}}
\newcommand{\vrad}{\ensuremath{v_\mathrm{rad}}}
\newcommand{\rg}{\ensuremath{r_\mathrm{g}}}
\newcommand{\rb}{\ensuremath{r_\mathrm{b}}}
\newcommand{\mbh}{\ensuremath{M_\mathrm{BH}}}
\newcommand{\msun}{\ensuremath{M_\odot}}
\newcommand{\lsun}{\ensuremath{L_\odot}}
\newcommand{\sigmaturb}{\ensuremath{\sigma_\mathrm{turb}}}

\newcommand{\vc}{\ensuremath{v_\mathrm{c}}}
\newcommand{\per}{\ensuremath{^{-1}}}
\newcommand{\pertwo}{\ensuremath{^{-2}}}
\newcommand{\rfit}{\ensuremath{r_\mathrm{fit}}}
\newcommand{\chisq}{\ensuremath{\chi^2}}
\newcommand{\chisqnu}{\ensuremath{\chi^2_\nu}}
\newcommand{\ndof}{\ensuremath{N_\mathrm{dof}}}

\newcommand{\upsh}{\ensuremath{\Upsilon_H}}
\newcommand{\upsj}{\ensuremath{\Upsilon_J}}

\newcommand{\cotwo}{CO(2$-$1)}
\newcommand{\cothree}{CO(3$-$2)}
\newcommand{\csfive}{CS(5$-$4)}

\begin{document}

\title{Black Hole Mass Measurements of Radio Galaxies NGC 315 and \\NGC 4261 Using ALMA CO Observations\footnote{Based on observations made with the NASA/ESA Hubble Space Telescope, obtained at the Space Telescope Science Institute, which is operated by the Association of Universities for Research in Astronomy, Inc., under NASA contract NAS 5-26555. These observations are associated with programs \#5124, 6673, 14219, and 15909.}}

\author[0000-0001-6301-570X]{Benjamin D. Boizelle}
\affiliation{Department of Physics and Astronomy, N284 ESC, Brigham Young University, Provo, UT, 84602, USA}
\affiliation{George P. and Cynthia Woods Mitchell Institute for Fundamental Physics and Astronomy, 4242 TAMU, Texas A\&M University, College Station, TX, 77843-4242, USA}
\email{boizellb@byu.edu}

\author[0000-0002-1881-5908]{Jonelle L. Walsh}
\affiliation{George P. and Cynthia Woods Mitchell Institute for Fundamental Physics and Astronomy, 4242 TAMU, Texas A\&M University, College Station, TX, 77843-4242, USA}

\author[0000-0002-3026-0562]{Aaron J. Barth}
\affiliation{Department of Physics and Astronomy, 4129 Frederick Reines Hall, University of California, Irvine, CA, 92697-4575, USA}

\author[0000-0002-3202-9487]{David A. Buote}
\affiliation{Department of Physics and Astronomy, 4129 Frederick Reines Hall, University of California, Irvine, CA, 92697-4575, USA}

\author[0000-0002-7892-396X]{Andrew J. Baker}
\affiliation{Department of Physics and Astronomy, Rutgers, the State University of New Jersey, 136 Frelinghuysen Road Piscataway, NJ 08854-8019, USA}

\author[0000-0003-2511-2060]{Jeremy Darling}
\affiliation{Center for Astrophysics and Space Astronomy, Department of Astrophysical and Planetary Sciences, University of Colorado, 389 UCB, Boulder, CO 80309-0389, USA}

\author[0000-0001-6947-5846]{Luis C. Ho}
\affiliation{Kavli Institute for Astronomy and Astrophysics, Peking University, Beijing 100871, China; Department of Astronomy, School of Physics, Peking University, Beijing 100871, China}

\author[0000-0003-1420-6037]{Jonathan Cohn}
\affiliation{George P. and Cynthia Woods Mitchell Institute for Fundamental Physics and Astronomy, 4242 TAMU, Texas A\&M University, College Station, TX, 77843-4242, USA}

\author[0000-0003-2632-8875]{Kyle M. Kabasares}
\affiliation{Department of Physics and Astronomy, 4129 Frederick Reines Hall, University of California, Irvine, CA, 92697-4575, USA}

\begin{abstract}
We present Atacama Large Millimeter/submillimeter Array (ALMA) Cycle 5 and Cycle 6 observations of \cotwo\ and \cothree\ emission at $0\farcs2-0\farcs3$ resolution in two radio-bright, brightest group/cluster early-type galaxies, NGC 315 and NGC 4261. The data resolve CO emission that extends within their black hole (BH) spheres of influence (\rg), tracing regular Keplerian rotation down to just tens of parsecs from the BHs. The projected molecular gas speeds in the highly inclined ($i\gtrsim60\degr$) disks rises at least 500 \kms\ near their galaxy centers. We fit dynamical models of thin-disk rotation directly to the ALMA data cubes, and account for the extended stellar mass distributions by constructing galaxy surface brightness profiles corrected for a range of plausible dust extinction values. The best-fit models yield $(\mbh/10^9\,\msun)=2.08\pm0.01(\mathrm{stat})^{+0.32}_{-0.14}(\mathrm{sys})$ for NGC 315 and $(\mbh/10^9\,\msun)=1.67\pm0.10(\mathrm{stat})^{+0.39}_{-0.24}(\mathrm{sys})$ for NGC 4261, the latter of which is larger than previous estimates by a factor of $\sim$3. The BH masses are broadly consistent with the relations between BH masses and host galaxy properties. These are among the first ALMA observations to map dynamically cold gas kinematics well within the BH-dominated regions of radio galaxies, resolving the respective \rg\ by factors of {$\sim$}5$-$10. The observations demonstrate ALMA's ability to precisely measure BH masses in active galaxies, which will enable more confident probes of accretion physics for the most massive galaxies.
\end{abstract}



\section{Introduction\label{sec:intro}}

At the center of presumably every large galaxy resides a supermassive black hole (BH), and dynamical measurements of the BH masses (\mbh) of over 100 galaxies have been made in the past two decades \citep[e.g.,][and references therein]{kormendy13,saglia16}. As a result, strong correlations have been established between the BH mass and several large-scale host galaxy properties, especially the stellar velocity dispersion \citep[$\sigma_\star$; e.g.,][]{gebhardt00,ferrarese00} and the stellar bulge luminosity \citep[$L_\mathrm{bul}$; e.g.,][]{kormendy95} and mass \citep[$M_\mathrm{bul}$; e.g.,][]{mcconnell13a}. These empirical relations suggest that the central BH and host galaxy grow in tandem through a series of accretion and merger events with feedback that regulates star formation.

The current data hint at a steeper $\mbh-\sigma_\star$ slope for the most luminous early-type galaxies \citep[ETGs; see][]{lauer07,bernardi07}, which include several brightest group/cluster galaxies (BGGs/BCGs). One plausible explanation is that BH growth follows a different evolutionary track in merger-rich environments \citep[e.g., see][]{bogdan18}. However, the BH census remains incomplete above $\sim$10$^9$ \msun, and typical \mbh\ uncertainties are of order 25\% in this high-mass regime \citep[e.g.,][]{saglia16}. In addition, potentially serious, and often unexplored, systematics in both stellar and gas-dynamical models may bias \mbh\ determinations \citep[for a discussion, see][]{kormendy13}. Any confident interpretation of BGG/BCG BH growth will require both larger numbers of \mbh\ measurements and greater measurement precision.

The most reliable BH mass measurements come from spatially resolved tracer kinematics that originate well within the sphere of influence ($\rg\approx G\mbh/\sigma_\star^2$). Within this radius, the BH's gravitational influence dominates over the galaxy's extended mass contributions. Circumnuclear gas disks have long been appealing kinematic targets for constraining BH masses because they are insensitive to large-scale galaxy properties, although high gas turbulence in many cases and potential non-circular motion have limited the usefulness of ionized gas tracers \citep{verdoes06}. Molecular gas disks, likely to have significantly less turbulent motion, offer an attractive alternative to ionized gas. In addition, very long baseline interferometry of water megamaser disks reveal emission arising from very deep within \rg\ \citep[e.g.,][]{kuo11,zhao18}. However, megamaser disks are rare and tend to be found in late-type galaxies with $\mbh\sim10^6-10^7$ \msun. Searches for maser emission in the most massive galaxies have so far been unsuccessful \citep{vandenbosch16a}. \citet{davis13} demonstrated that the more common tracer CO could be used to constrain BH masses in ETGs, and the increased sensitivity and angular resolution of the Atacama Large Millimeter/submillimeter Array (ALMA) relative to the previous generation of mm/sub-mm facilities have reinvigorated gas-dynamical efforts.

About 10\% of all luminous ETGs exhibit morphologically round dust features in Hubble Space Telescope (\hst) images \citep[e.g.,][]{lauer05}, and such dust is often associated with regular molecular gas kinematics \citep{alatalo13}. We therefore began an ALMA campaign to map \cotwo\ kinematics in nearby ETGs with targets selected based on dust morphology. With our ALMA Cycle 2 sample \citep[][hereinafter Paper I]{boizelle17}, we demonstrated that the molecular gas disks were both dynamically cold (with the ratio of intrinsic line width to circular speed $\ll1$) and typically only mildly warped. The disks can be nearly ideal probes of the central gravitational potential and are very amenable to detailed gas-dynamical modeling. \citet{barth16b} first demonstrated ALMA's power to constrain BH masses, and follow-up CO imaging at high angular resolution for two targets in our Cycle 2 sample has enabled \mbh\ determinations with percent-level BH mass precision \citep[][hereinafter Paper II]{barth16a,boizelle19}. Other groups have simultaneously pioneered this new avenue for measuring BH masses in similar targets \citep{onishi15,onishi17,davis17b,davis18,smith19,nagai19,north19,vila19} with typical $\sim$10$-$20\% precision, and the method has been applied to spiral galaxies \citep{nguyen20} and dwarf elliptical galaxies \citep{davis20} as well.

As part of our ALMA program, we selected NGC 315 and NGC 4261 for \cotwo\ observations after identifying 2\arcsec\ to 3\arcsec-wide, morphologically round dust disks in archival \hst\ imaging. Previous CO emission-line surveys reported a tentative CO detection for NGC 315, corresponding to $\log_{10}[M_\mathrm{H2}/\msun]\sim7.91$, while NGC 4261 was undetected, with an H$_2$ mass upper limit of $\log_{10}[M_\mathrm{H2}/\msun]\lesssim7.68$ \citep{combes07,ocana10,young11,davis19}. Nevertheless, we expected to easily detect and resolve the disks with ALMA. For comparison, there has been a $\log_{10}[M_\mathrm{H2}/\msun]\lesssim6.0-6.8$ threshold for similar, nearby ($\lesssim$100 Mpc) targets observed with ALMA (\citetalias{boizelle17}; \citealp{ruffa19a}). We expected both galaxies to fall at the upper end of the BH scaling relations given the stellar velocity dispersions and luminosities reported in the HyperLeda database \citep{makarov14}. Based on their large anticipated \rg, we imaged \cotwo\ in NGC 315 and NGC 4261 at $\sim$0\farcs3-resolution in order to at least partially resolve CO kinematics within the BH-dominated regions.

NGC 315 dominates a sparse group environment \citep{nothenius93,crook07} near the confluence of galaxy filaments in the Pisces-Perseus supercluster \citep{ensslin2001}. It is a Fanaroff-Riley (FR) Type I \citep{laing06} cD galaxy \citep{devaucouleurs91}. At its center resides a type 1.9 low-ionization nuclear emission region \citep[LINER;][]{ho97a}, with unresolved non-stellar emission detected at radio to X-ray wavelengths \citep{venturi93,worrall07,gu07}. Redshift-independent distances for the galaxy from the NASA/IPAC Extragalactic Database\footnote{\url{http://ned.ipac.caltech.edu}} range from $\sim$50$-$100 Mpc, and a forthcoming surface brightness fluctuation (SBF) distance measurement \citep{goullaud18} should clarify the uncertainty. Here, we adopt the Hubble flow distance assuming the Virgo $+$ Great Attractor $+$ Shapley Supercluster inflow model for a $\Lambda$CDM cosmology with $\Omega_\mathrm{matter}=0.308$, $\Omega_\mathrm{vacuum}=0.692$, and $H_0=67.8$ \kms\ Mpc\per\ \citep{planck16}. This assumption corresponds to a luminosity distance of 72.3 Mpc, an angular size distance of 70.0 Mpc, and an angular scale of 340 pc arcsec\per. The derived \mbh\ scales linearly with the assumed distance, so any change to the distance will result in a commensurate shift in \mbh. No previous study has constrained the NGC 315 BH mass.

\begin{figure*}[ht]
\begin{center}
\includegraphics[trim=0mm 0mm 0mm 0mm, clip, width=\textwidth]{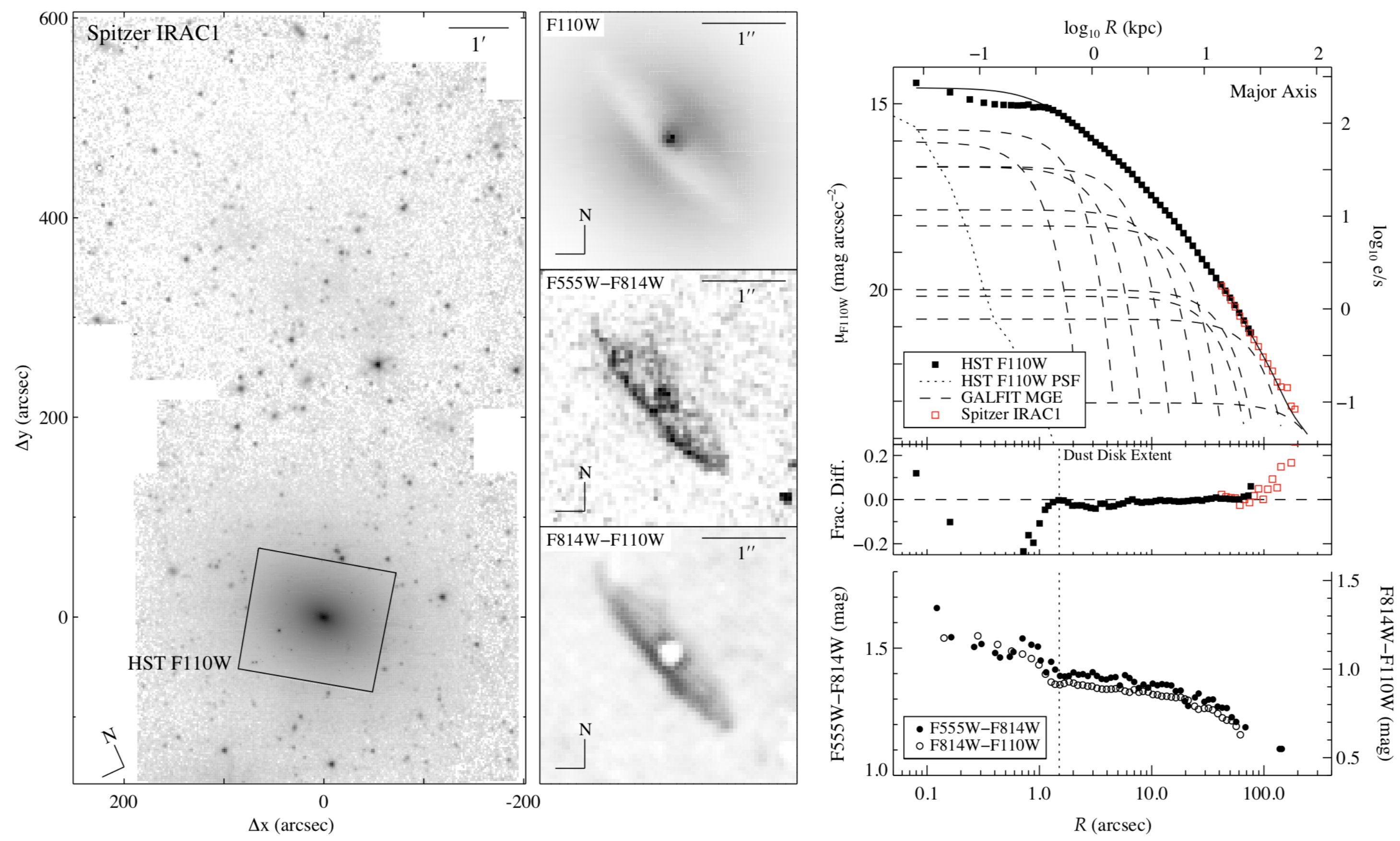}
\begin{singlespace}
  \caption{Optical to mid-IR imaging of NGC 315. The \hst\ WFC3 F110W image is overlaid on a \spitzer\ IRAC1 mosaic (\textit{left panel}), shown using a logarithmic intensity scale. F555W$-$F814W and F814W$-$F110W color maps (shown in central $4\farcs1\times 4\farcs1$ cut-out regions; \textit{middle panels}) trace a morphologically regular and highly inclined dust disk. Luminous mass models for this galaxy were constructed by parameterizing the $J$-band stellar surface brightness using a \texttt{GALFIT} MGE (\textit{top right panel}) accompanied by a nuclear point-source component. Here, we augmented the \hst\ measurements with larger-scale IRAC1 data. The ($A_J=0$ mag) MGE solution shown here and in Table~\ref{tbl:tbl1} was derived after masking the most dust-obscured regions seen in the $J$-band cut-out.\label{fig:fig1}}
\end{singlespace}
\end{center}
\end{figure*}

NGC 4261 (3C 270) is the brightest galaxy in the NGC 4261 group \citep{davis95} in the direction of the Virgo cluster. This FR I \citep{jaffe94} E2-3 galaxy \citep{devaucouleurs91} has a distance modulus from a SBF measurement of $m-M=32.50\pm0.19$ mag \citep{tonry01}, which translates to a luminosity distance of $31.6\pm2.8$ Mpc. Using the observed redshift of $z=0.00746$ \citep{huchra12}, the angular size distance is 31.1 Mpc, and we adopted an angular scale of 150.9 pc arcsec\per. Two previous studies estimated a BH mass of $\sim5\times10^8$ \msun\ for NGC 4261 \citep{ferrarese96,humphrey09}.

In this paper, we present the first ALMA 12-m CO imaging of NGC 315 and NGC 4261. While the focus remains on our $\sim$0\farcs3-resolution \cotwo\ observations, we also include additional \cothree\ imaging of NGC 4261 from the ALMA archive to test the robustness of the \cotwo\ gas-dynamical modeling results. The paper is organized as follows. In Section~\ref{sec:oir}, we describe the \hst\ and \spitzer\ Space Telescope observations used to measure the galaxies' stellar surface brightness distributions. In Section~\ref{sec:almadata}, we introduce the ALMA CO imaging. We summarize our gas-dynamical modeling method and results in Section~\ref{sec:dynmod}, and include an examination of various sources of uncertainty in the \mbh\ determination. In Section~\ref{sec:discussion}, we compare to the BH$-$galaxy scaling relations, discuss future improvements to the ALMA-based \mbh\ measurements, and examine prospects for precision BH mass determination in galaxies with mm/sub-mm-bright nuclei. We conclude in Section~\ref{sec:conclusion}.

\section{Optical and Infrared Observations\label{sec:oir}}

We used \hst\ Wide Field Camera 3 \citep[WFC3;][]{dressel19} infrared (IR) data to determine luminous mass distributions for NGC 315 and NGC 4261, and WFC3/IR and Wide Field Planetary Camera 2 \citep[WFPC2;][]{holtzman95} optical imaging to quantify the potential impact of dust on our luminous mass models. In order to accurately constrain the stellar halos in these galaxies, we supplemented the \hst\ data with large-scale \spitzer\ InfraRed Array Camera \citep[IRAC;][]{fazio04} observations. Below, we summarize the observations, present surface brightness measurements, and explain dust extinction models.

\begin{figure*}[ht]
\begin{center}
\includegraphics[trim=0mm 0mm 0mm 0mm, clip, width=\textwidth]{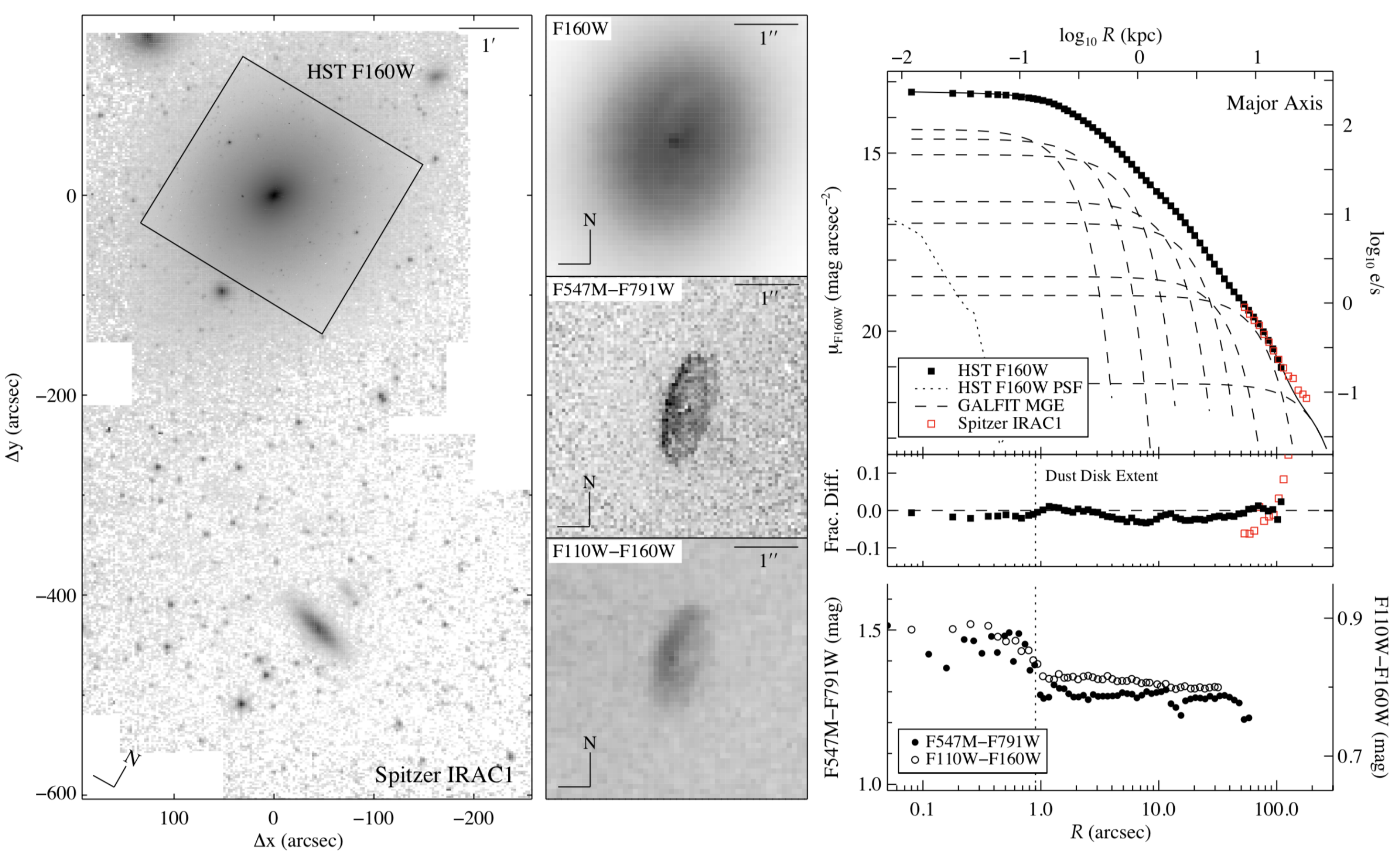}
\begin{singlespace}
  \caption{Optical to mid-IR imaging of NGC 4261. The \hst\ WFC3 F160W image is overlaid on a \spitzer\ IRAC1 mosaic (\textit{left panel}), shown using a logarithmic intensity scale. The F547M$-$F791W and F110W$-$F160W color maps (shown in central $4\farcs1\times 4\farcs1$ cut-out regions; \textit{middle panels}) reveal a round circumnuclear dust disk that is slightly miscentered relative to the stellar bulge. The luminous mass model for this galaxy was constructed by parameterizing the $H$-band surface brightness using a \texttt{GALFIT} MGE and a central point source component (\textit{top right panel}). These \hst\ measurements were augmented with larger-scale IRAC1 data. The ($A_H=0$ mag) MGE solution shown here and in Table~\ref{fig:fig1} was derived after masking the reddest IR colors in the $H$-band cutout, which are expected to trace the most dust-obscured regions.\label{fig:fig2}}
\end{singlespace}
\end{center}
\end{figure*}

\begin{deluxetable*}{cccccccc}[ht]
\tabletypesize{\small}
\tablecaption{MGE Parameters\label{tbl:tbl1}}
\tablewidth{0pt}
\tablehead{
\colhead{$j$} & 
\colhead{$\log_{10}$ $I_{J,j}$ (\lsun\ pc\pertwo)} & 
\colhead{$\sigma_{j}^\prime{}$ (arcsec)} & 
\colhead{$q_{j}^\prime{}$} &
 &
\colhead{$\log_{10}$ $I_{H,j}$ (\lsun\ pc\pertwo)} & 
\colhead{$\sigma_{j}^\prime{}$ (arcsec)} & 
\colhead{$q_{j}^\prime{}$}
\\[-1.5ex]
\colhead{(1)} & 
\colhead{(2)} & 
\colhead{(3)} & 
\colhead{(4)} &
 &
\colhead{(2)} & 
\colhead{(3)} & 
\colhead{(4)}
}
\startdata
  & \multicolumn{3}{c}{\bf NGC 315} &  & \multicolumn{3}{c}{\bf NGC 4261} \\ \cline{2-4} \cline{6-8}
1 & 3.798 & 0.580 & 0.871 &  & 4.261 & 1.075 & 0.830 \\
2 & 3.895 & 1.237 & 0.786 &  & 4.143 & 2.131 & 0.717 \\
3 & 3.485 & 2.347 & 0.704 &  & 3.964 & 3.837 & 0.729 \\
4 & 3.483 & 4.132 & 0.722 &  & 3.437 & 8.191 & 0.719 \\
5 & 3.017 & 8.191 & 0.664 &  & 3.194 & 13.54 & 0.834 \\
6 & 2.844 & 13.25 & 0.748 &  & 2.595 & 23.79 & 0.816 \\
7 & 2.085 & 26.51 & 0.763 &  & 2.381 & 49.05 & 0.862 \\
8 & 2.155 & 30.90 & 0.689 &  & \phantom{*}1.392* & \phantom{*}144.4* & \phantom{*}0.820* \\
9 & \phantom{*}1.839* & \phantom{*}61.95* & \phantom{*}0.810* &  & \nodata & \nodata & \nodata \\
10 & \phantom{*}0.939* & \phantom{*}192.6* & \phantom{*}0.980* &  & \nodata & \nodata & \nodata
\enddata
\begin{singlespace}
  \tablecomments{NGC 315 and NGC 4261 MGE solutions constructed from their respective $J$ and $H$-band mosaics. For NGC 315, the MGE has a uniform $\mathrm{PA}=44.31\degr$ for all components, while $\mathrm{PA}=-22.03\degr$ for the best-fit NGC 4261 model. Column (1) lists the component number, column (2) is the central surface brightness assuming absolute solar magnitudes of $M_{\odot,\,J}=3.82$ mag and $M_{\odot,\,H}=3.37$ mag \citep{willmer18}, column (3) gives the Gaussian standard deviation along the major axis, and column (4) provides the component axis ratio. Primes indicate projected quantities. \spitzer\ IRAC1 MGE components identified with an asterisk were scaled to match the WFC3/IR data and included as fixed components when fitting MGEs to the \hst\ images. The unresolved nuclear features were modeled as point sources during the MGE fit with \texttt{GALFIT}, and have apparent magnitudes of $m_J\approx19.2$ mag and $m_H\approx19.6$ mag for NGC 315 and NGC 4261, respectively.}
\end{singlespace}
\end{deluxetable*}

\subsection{\hst\ Imaging\label{sec:hstimaging}}

For NGC 315, we retrieved archival WFC3/IR data \citep[GO-14219;][]{goullaud18} taken in the F110W filter (hereinafter $J$-band), which covers the inner $2\farcm1\times 2\farcm2$ region of the galaxy. We combined calibrated $J$-band frames using \texttt{AstroDrizzle} \citep{gonzaga12} to produce a final mosaic with a 0\farcs08 pixel\per\ scale. We did not remove the background level during the drizzling process. In Figure~\ref{fig:fig1}, we show the final $J$-band image that probes the stellar surface brightness out to $R\sim100\arcsec$ ($\sim$34 kpc). The highly inclined dust disk noticeably suppresses stellar light even in the $J$-band image.

For NGC 4261, we obtained new WFC3/IR data (GO-15909; PI: Boizelle) taken in the F160W filter (hereinafter $H$-band) and the $J$-band filter. For the $H$-band observations, we used a large-scale mosaic pattern to probe the galaxy light out to $R\sim150\arcsec$ ($\sim$23 kpc) and short subarray exposures taken in a 4-point dither pattern centered on the nucleus to sample the point-spread function (PSF) well. In contrast, the $J$-band observations consisted of 4-point and 2-point dithered subarray exposures covering a $\sim 1\farcs2 \times 1\farcs0$ field. We created composite images with a pixel scale of 0\farcs08 pixel\per\ for each filter and sequence using \texttt{AstroDrizzle}. To construct a final $H$-band image, we replaced the central portions of the drizzled large-scale mosaic with the subarray data. Except for a region to the northeast on the dust disk's near side (see Figure~\ref{fig:fig2}), the $H$-band data appear largely unaffected by circumnuclear dust.

In order to characterize the dust attenuation properties in the two galaxies (see Section~\ref{sec:extinction}), we obtained additional optical data from the \hst\ archive. For NGC 315, we retrieved WFPC2 F555W and F814W images \citep[GO-6673;][]{verdoes99} and aligned these data to the $J$-band mosaic using \texttt{TweakReg} \citep{gonzaga12}. To take advantage of the better angular resolution, we drizzled the WFPC2 data to a 0\farcs05 pixel\per\ scale. For NGC 4261, we retrieved WFPC2 data taken in the F547M and F791W filters \citep[GO-5124;][]{ferrarese96}, which avoid prominent emission lines from the active galactic nucleus (AGN). These data were likewise drizzled to 0\farcs05 pixel\per\ scales before being manually aligned to the $H$-band mosaic. As demonstrated in Figures~\ref{fig:fig1} and \ref{fig:fig2}, the observed optical color gradients for NGC 315 and NGC 4261 are consistent with those measured for other massive, red ETGs \citep[e.g.,][]{labarbera10,kennedy16}.

\subsection{\spitzer\ Imaging\label{sec:spitzerimaging}}

For both NGC 315 and NGC 4261, we retrieved IRAC channel 1 (3.6 \micron) data from the \spitzer\ Enhanced Imaging Products archive\footnote{\url{http://irsa.ipac.caltech.edu/data/SPITZER/Enhanced/SEIP/overview.html}}. In addition to sensitivity considerations, we selected this channel to trace the old stellar population while avoiding emission-line contamination in redder filters. The IRAC1 super-mosaics shown in Figures~\ref{fig:fig1} and \ref{fig:fig2} cover both galaxies out to $R\sim11\arcmin$ ($\sim$225 and 100 kpc for NGC 315 and NGC 4261, respectively). We measured and removed any residual background from the \spitzer\ data using uncontaminated regions far from the primary targets.

\subsection{Stellar Luminosity Profile\label{sec:stellarprofile}}

The NGC 315 and NGC 4261 stellar halos clearly extend beyond the \hst\ IR mosaics. Following the method outlined in \citetalias{boizelle19}, we used large-scale \spitzer\ imaging to constrain IR zodiacal background levels \citep{pirzkal14} and to extend stellar surface brightness measurements. After masking galaxies and foreground stars, we measured WFC3/IR and IRAC1 surface brightness profiles for NGC 315 and NGC 4261 in directions toward the outermost edges of the 3.6 $\micron$ super-mosaics. Using overlapping measurements between $R\sim 50-100\arcsec$, we simultaneously determined a background level of $\mu_J=21.68$ mag arcsec\pertwo\ and a color of $J-\mathrm{IRAC1}=1.53$ mag needed to align the NGC 315 measurements. Likewise, we determined a background level of $\mu_H=20.44$ mag arcsec\pertwo\ with $H-\mathrm{IRAC1}=0.72$ mag for NGC 4261. Nearly identical values were obtained when matching \hst\ and \spitzer\ surface brightness profiles along different directions.

To characterize the stellar luminosity distributions, we modeled both the WFC3/IR mosaics and flux-scaled IRAC1 images with concentric 2D Gaussian functions using \texttt{GALFIT} \citep{peng02}. Preliminary parameter guesses were obtained using the \citet{cappellari02} Multi-Gaussian Expansion (MGE) code, and the major-axis position angles (PAs) of the Gaussian components were tied together. We adopted Tiny Tim models \citep{krist04} that were dithered and drizzled identically to the \hst\ observations to account for PSF blurring when modeling the WFC3/IR data. For the IRAC1 observations, we employed an empirical model of the \spitzer\ point response function \citep{hoffman04}.

We started by fitting MGEs to the WFC3 and IRAC1 images separately. The largest \spitzer\ MGE full-widths at half-maxima (FWHMs) were at least twice that of the outermost \hst\ MGE component, so we incorporated the outermost IRAC1 Gaussians into the initial \hst\ MGE. We then re-optimized the MGE fits to the WFC3 mosaics only, holding fixed the (scaled) magnitudes and FWHMs of the extended IRAC1 Gaussian components while tying their centroids and PAs to those of the inner \hst\ Gaussian components. Such large-scale information is not necessary for the gas-dynamical models described here, but these MGEs may be useful for future studies of the galaxies (e.g., stellar-dynamical modeling).

\subsubsection{Application to NGC 315\label{sec:stellarprofilesub1}}

Circumnuclear dust significantly obscures the central stellar light of NGC 315, even at the WFC3/IR wavelengths. The impact of dust is explored more in Section~\ref{sec:extinction}, but we constructed the first of our $J$-band MGEs after simply masking the most dust-obscured pixels where $I-J>0.96$ mag. The mask excludes filamentary dust features to the south and west of the nucleus and the entire disk region except for the central few pixels and those behind (to the northwest of) the galaxy center. Initial \texttt{GALFIT} fits preferred the innermost Gaussian to be very compact (FWHM < 0\farcs01), consistent with unresolved emission from the AGN. To account for non-stellar contributions in the \texttt{GALFIT} fit, we included a concentric PSF component that follows the F110W Tiny Tim model.

By comparing the data and MGE model, we estimate that the intrinsic disk extinction may reach as high as $A_J\sim1.4$ mag at a semi-major axis distance of $R\sim0\farcs5$. This MGE solution (shown in Figure~\ref{fig:fig1} and in Table~\ref{tbl:tbl1}) consists of 10 Gaussian components, with the outermost two Gaussians based on the large-scale \spitzer\ surface brightness. Each of the Gaussian magnitudes was corrected for Galactic reddening, assuming $A_{J,\mathrm{Gal}}=0.057$ mag towards NGC 315 \citep{schlafly11}. The model's $J$-band luminosity measured within the central 300\arcsec\ ($\sim$100 kpc) is $L_J=5.79\times10^{11}$ \lsun, and we estimate a circularized half-light radius \citep{cappellari13a} of $R_e=56\farcs0$ ($\sim$19.0 kpc) in this filter, roughly a factor of two larger than curve-of-growth estimates \citep{vandenbosch16b,veale17}.

\begin{figure}[ht]
\begin{center}
\includegraphics[trim=0mm 0mm 0mm 0mm, clip, width=\columnwidth]{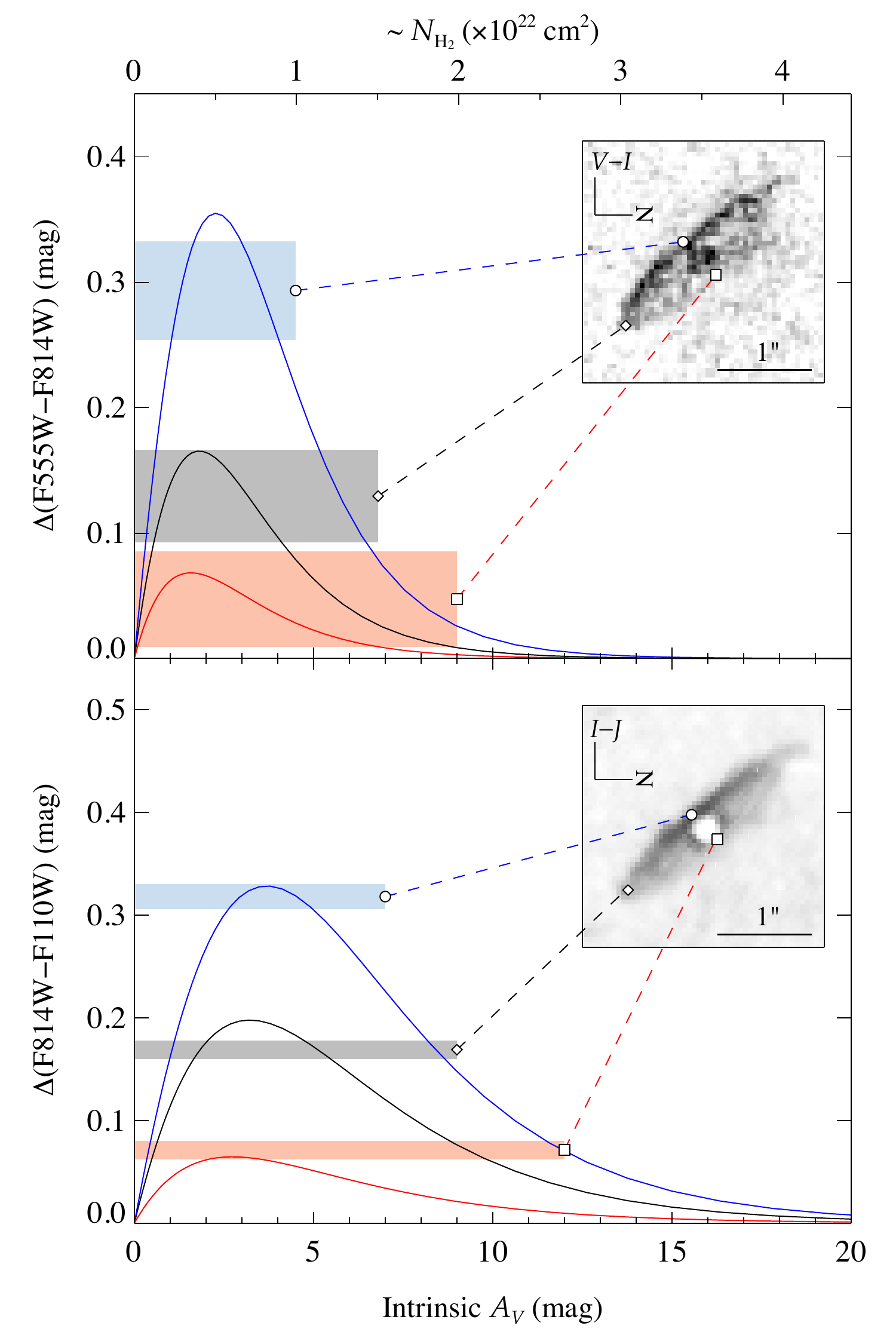}
\begin{singlespace}
  \caption{Comparison between observed color excesses at the marked locations on the NGC 315 \hst\ F555W$-$F814W and F814W$-$F110W maps and the curves expected for a simple dust attenuation model. Shaded rectangles encompass the range of observed color values about the marked spatial locations. Agreement between observations and the model along the ring of maximal observed color suggests the stellar light behind the galaxy's midplane is obscured by dust reaching an intrinsic $A_V\sim3-4$ mag ($A_J\sim 1-1.5$ mag) at a projected separation of $R\sim0\farcs2$ from the nucleus along the minor axis. H$_2$ column densities were estimated using a Galactic $N_{\mathrm{H2}_2}/A_V$ ratio \citep{guver09}.\label{fig:fig3}}
\end{singlespace}
\end{center}
\end{figure}

\subsubsection{Application to NGC 4261\label{sec:stellarprofilesub2}}

While much less apparent than for NGC 315, circumnuclear dust still reduces the central surface brightness of the NGC 4261 WFC3/IR data. We started by masking the most likely dust-obscured regions, excluding pixels where $J-H>0.85$ mag to the south of the nucleus and along the disk's eastern (near) side. We then parameterized the $H$-band data with a \texttt{GALFIT} MGE and a PSF component. In the same manner as done for NGC 315, we included the single most extended \spitzer\ MGE component as a fixed Gaussian during the fit to the \hst\ data. This MGE solution, shown in Figure~\ref{fig:fig2} and in Table~\ref{tbl:tbl1}, has eight Gaussian components, each of which was corrected for Galactic reddening assuming $A_{H,\mathrm{Gal}}=0.009$ mag towards NGC 4261 \citep{schlafly11}. The model's total $H$-band luminosity measured within the central 300\arcsec\ ($\sim$45 kpc) is $L_H=2.29\times10^{11}$ \lsun, and we found $R_e=36\farcs2$ ($\sim$5.5 kpc), which is in good agreement with an average of early curve-of-growth estimates \citep{cappellari11} but is roughly 30\% smaller than more recent values \citep{vandenbosch16b,cappellari13a}.

\begin{figure*}[ht]
\begin{center}
\includegraphics[trim=0mm 0mm 0mm 0mm, clip, width=\textwidth]{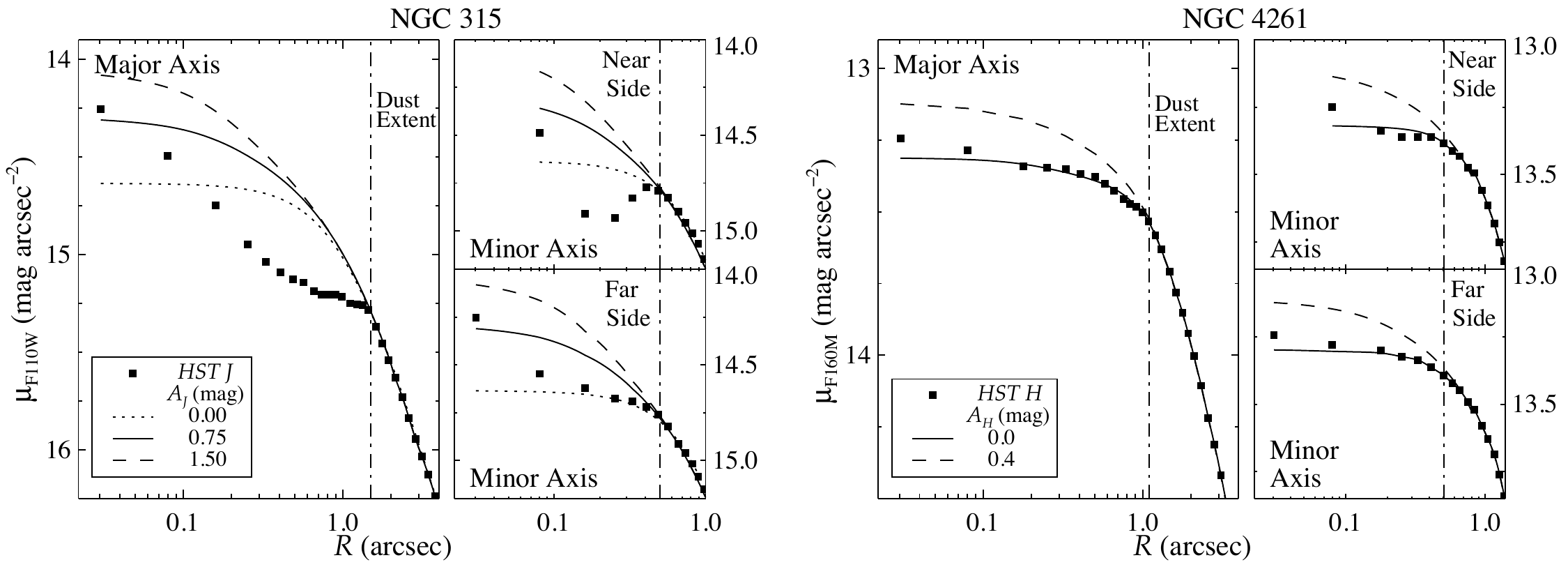}
\begin{singlespace}
  \caption{Comparison between the observed NGC 315 $J$ and NGC 4261 $H$-band surface brightness profiles and those derived from extinction-corrected MGE models. For NGC 315, we show both the dust-masked ($A_J=0$ mag) model and MGEs constructed after correcting central surface brightness values for $A_J=0.75$ and 1.50 mag of extinction of the light behind the disk. For NGC 4261, we show the dust-masked ($A_H=0$ mag) model along with an MGE solution corrected for a central extinction of $A_H=0.4$ mag. The minor axis plots (\textit{right-hand panels}) show NGC 315 and NGC 4261 surface brightness profiles taken along the near (southeast and east of the nuclei, respectively) and far sides of the dust disks. We note that the central $J$ and $H$-band excesses are fit as point sources with no extinction correction and are not included in the MGE models.\label{fig:fig4}}
\end{singlespace}
\end{center}
\end{figure*}

\subsection{Disk Extinction Modeling\label{sec:extinction}}

Based on our previous analysis in \citetalias{boizelle19}, even a modest change in the central stellar contribution to the gravitational potential can produce a shift in the best-fit \mbh\ that far exceeds both the statistical uncertainty from dynamical modeling and most other systematic error terms. Therefore, using \hst\ color maps, we determined a plausible range of central extinction values, following the method outlined here and described in greater detail in \citetalias{boizelle19} \citep[see also][]{goudfrooij94,viaene17}. 

Our method assumes that a geometrically thin, inclined dust disk lies in the midplane of an oblate axisymmetric galaxy. If the dust disk has a very low optical depth, then we expect there to be little to no color excess. The same is true if the disk has a very high optical depth because the starlight behind the disk is completely obscured. For some moderate value of disk optical depth, the color excess should be large, as some starlight passes through the disk but becomes reddened. Therefore, the color excess should increase, reach a maximum value, and then decrease as the disk optical depth increases. In addition, we expect there to be a spatial dependence due to the inclined nature of the disk. For a particular value of the disk optical depth, the near side of the disk should have a larger color excess compared to the far side of the disk because there is a larger fraction of starlight behind the near side of the disk. We indeed see such behavior when employing a simple embedded screen model, as discussed below.

After deprojecting the NGC 315 MGE in Table~\ref{tbl:tbl1} for an inclination angle $i=75\degr$ (inferred from our initial gas-dynamical models), we estimated the fraction of total stellar light originating from behind the midplane on a pixel-by-pixel basis. Adopting a ratio of total-to-selective extinction of $R_V=3.1$ \citep{mathis90}, we determined model color excess curves as a function of the intrinsic extinction $A_V$ of the obscuring disk. Comparing the model color excess curves with the data (Figure~\ref{fig:fig3}) suggests extinction values of $A_V\sim3-4$ mag ($A_J\sim1-1.5$ mag) at a projected $R\sim0\farcs2$ from the nucleus along the minor axis.

In addition to the initial ($A_J=0$ mag) MGE solution presented in Figure~\ref{fig:fig1} and Table~\ref{tbl:tbl1}, we constructed two other luminous mass models for NGC 315 to cover the plausible range in $A_J$. To start, we masked nearly the entire dust disk as described in Section~\ref{sec:stellarprofilesub1} before fitting the $J$-band mosaic for $R<10\arcsec$ using a PSF-convolved Nuker function \citep{lauer95} and a concentric PSF component in \texttt{GALFIT}. The best-fit model has inner and outer power-law slopes of $\gamma=0.0$ and $\beta=1.48$, respectively, with a break radius $\rb=1\farcs42$ and a transition sharpness $\alpha=1.54$. After subtracting the point-source component, we corrected the central unmasked surface brightness values for extinction levels of $A_J=0.75$ and $1.50$ mag. Next, we refit surface brightnesses corrected for each $A_J$ value in turn with a Nuker function while fixing $\beta=1.48$ and $\rb=1\farcs42$. For the $A_J=0.75$ and $1.50$ mag cases, we found $\gamma=0.18$ and $0.33$, respectively. The best-fit $\gamma$ for the $A_J=1.50$ mag case is at the boundary between those of core and power-law galaxies \citep[e.g.,][]{faber97}. To construct ``dust-corrected'' luminous mass models, we replaced the $J$-band data within the disk region with the best-fit Nuker models. We optimized MGE models with \texttt{GALFIT} to the ``corrected'' images in nearly the same manner as before, although we did not apply a dust mask. The ``dust-corrected'' MGEs are given in Table~\ref{tbl:tbla} in the Appendix, and Figure~\ref{fig:fig4} compares the data and the three model surface brightness profiles.

The NGC 4261 $H$-band mosaic is not as visibly affected by dust as are the NGC 315 WFC3/IR data. To estimate the magnitude of the obscuration, we applied the same dust modeling technique used for NGC 315. Adopting $i=64\degr$ from fits to the dust disk morphology \citep{ferrarese96}, we deprojected the NGC 4261 initial ($A_H=0$ mag) MGE presented in Figure~\ref{fig:fig2} and Table~\ref{tbl:tbl1}, and compared the model and observed \textit{J}$-$\textit{H} color excesses. The maximal observed color excesses are consistent with $A_H\lesssim0.4$ mag at all radii. We then constructed a ``dust-corrected" MGE, first masking regions of the dust disk with high color excess [$\Delta(J-H)>0.05$] before fitting the $H$-band mosaic within $R<10\arcsec$ using a PSF-convolved Nuker function and a concentric PSF component. The best-fit Nuker model has power-law slopes of $\gamma=0.03$ and $\beta=1.43$, with $\rb=1\farcs72$ and $\alpha=2.47$. After subtracting the point-source component, we corrected the central surface brightness values for $A_H=0.4$ mag and fit the corresponding surface brightnesses using a Nuker function with fixed $\beta=1.43$ and $\rb=1\farcs72$, finding a best-fit inner slope of $\gamma=0.10$. We replaced the $H$-band surface brightness values within the dust disk region with the Nuker model and constructed a MGE that fit the ``corrected'' image without including a dust mask. We compare the data and model surface brightness profiles in Figure~\ref{fig:fig4} and present the ``dust-corrected" MGE in Table~\ref{tbl:tbla} in the Appendix.

\begin{figure*}[ht]
\begin{center}
\includegraphics[trim=0mm 0mm 0mm 0mm, clip, width=0.9\textwidth]{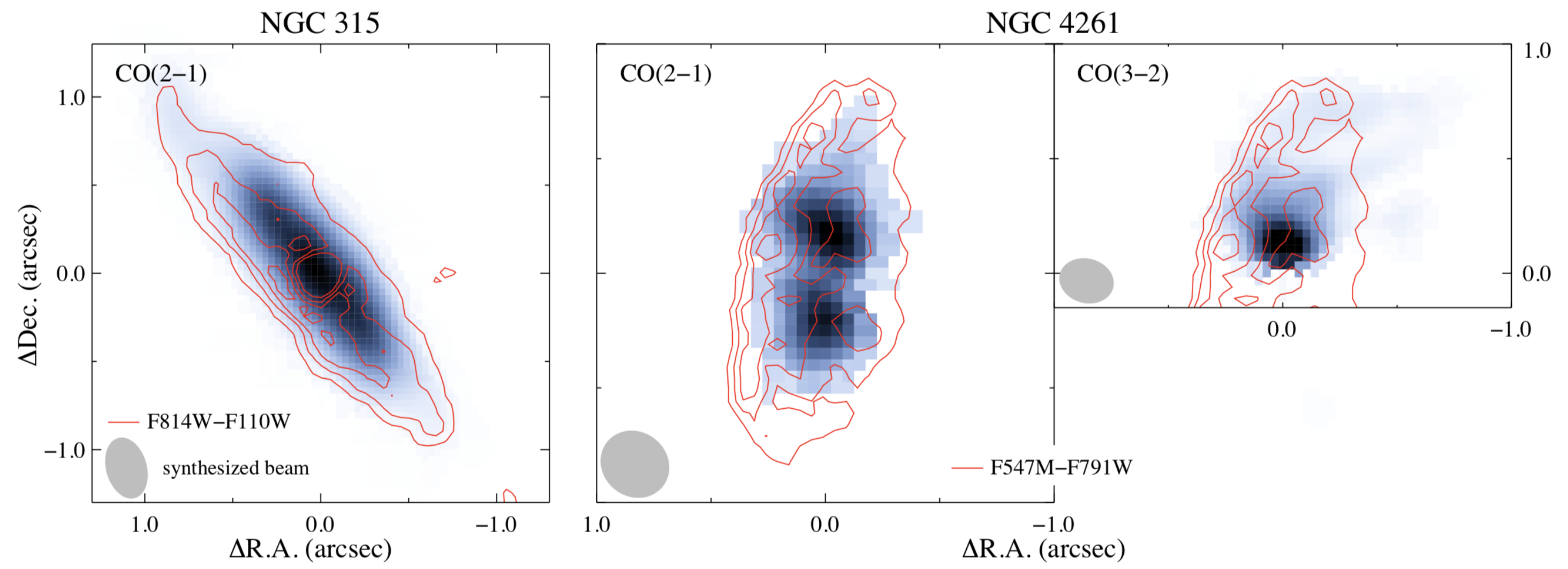}
\begin{singlespace}
  \caption{\cotwo\ and \cothree\ zeroth moment maps measured from the NGC 315 and NGC 4261 ALMA data sets. The frequency range of the archival NGC 4261 \cothree\ observations excludes most of the emission on the approaching (south) side of the disk. Contours show the respective \hst\ $I-J$ and $V-I$ colors. The CO emission in each galaxy is concentrated within the region enclosed by the maximum \hst\ color contours.\label{fig:fig5}}
\end{singlespace}
\end{center}
\end{figure*}

\section{ALMA Data\label{sec:almadata}}

\subsection{Observations and data processing\label{sec:almaobs}}

We obtained ALMA imaging of NGC 315 and NGC 4261 in the C43$-$5 configuration in Program 2017.1.00301.S. Observations consisted of a single pointing with four $\sim$2 GHz$-$bandwidth spectral windows: one was centered on the redshifted \cotwo\ 230.538 GHz line, two others measured the continuum at average sky frequencies of 228.4 and 243.0 GHz, and one was designed for the possible detection of redshifted \csfive\ 244.935 GHz line emission. For NGC 315, data were obtained during Cycle 6 in three Execution Blocks spanning 2018 October 19 $-$ November 19. Excluding data with a quality assurance rating of ``semi-pass'', the total on-source integration time was 89 minutes. For NGC 4261, our data were obtained during Cycle 5 with a single Execution Block on 2018 January 19 and had an on-source integration time of 31.5 minutes. The data were flux calibrated using ALMA quasar standards J2253$+$1608 and J1229$+$0203, which have absolute flux calibration uncertainties of $\sim$10\% at 1\,mm \citep{fomalont14}. We have propagated this systematic uncertainty into all subsequent flux and flux density measurements, and quantities derived from them.

Visibilities were calibrated using version 5.1.2 of the \texttt{Common Astronomy Software Applications} \citep[\texttt{CASA};][]{mcmullin07} package. We used the bright ($S_\mathrm{230\,GHz}=0.2-0.25$ Jy) continuum emission at the centers of NGC 315 and NGC 4261 to apply phase and amplitude self-calibration. Following $uv-$plane continuum subtraction, the \cotwo\ spectral window data were deconvolved using \texttt{CASA} \texttt{tclean} with Briggs \citep[][]{briggs95} weighting of $r=0.5$. While we detect resolved 1.3\,mm continuum in both galaxies and centrally concentrated CS(5$-$4) emission in NGC 315, we focus here only on the CO emission. For NGC 315, the synthesized beam at a sky frequency of $\sim$229 GHz has $\theta_\mathrm{FWHM}=0\farcs35 \times 0\farcs22$ oriented at $\mathrm{PA}=15.7\degr$. We imaged the primary spectral window into a \cotwo\ cube with $0\farcs035$ pixels and 7.81 MHz channels, corresponding to rest-frame velocity widths of about 10.3 \kms\ and resulting in typical root-mean-squared (rms) sensitivities of $\sim$0.35 mJy beam\per. At the same frequency for NGC 4261, imaging yields $\theta_\mathrm{FWHM}=0\farcs31 \times 0\farcs28$ at $\mathrm{PA}=51.9\degr$. Due to low S/N of the \cotwo\ line, we imaged NGC 4261 using coarser channels of 31.2 MHz (rest-frame 40.9 \kms), with an rms sensitivity of $\sim$0.26 mJy beam\per\ per channel. For the NGC 4261 \cotwo\ cube, we also selected a larger pixel size of 0\farcs05 pixel\per\ to appropriately sample $\theta_\mathrm{FWHM}$.

Just two days after our data were taken, ALMA observed NGC 4261 in Band 7 as a part of Program 2017.1.01638.S (PI: Kameno). One of the spectral windows had usable data extending from 341.75$-$343.45 GHz, which covers the \cothree\ line on the receding side of the disk, down to line-of-sight velocities (\vlos) of $-$158 \kms\ relative to the galaxy's systemic velocity (\vsys). We retrieved the Band 7 data to complement our \cotwo\ observations and reduced them using the same approaches to self-calibration and $uv-$plane continuum subtraction. We then imaged the visibilities into a \cothree\ data cube with $0\farcs035$ pixels and 15.6 MHz (rest-frame 13.8 \kms) channels with $\sim$0.34 mJy beam\per\ point-source sensitivities. At a sky frequency of $\sim$344 GHz, the Band 7 data have $\theta_\mathrm{FWHM}=0\farcs24 \times 0\farcs19$ at $\mathrm{PA}=75.2\degr$, resulting in 30\% better average linear resolution than our \cotwo\ observations.

\begin{figure*}[ht]
\begin{center}
\includegraphics[trim=0mm 0mm 0mm 0mm, clip, width=\textwidth]{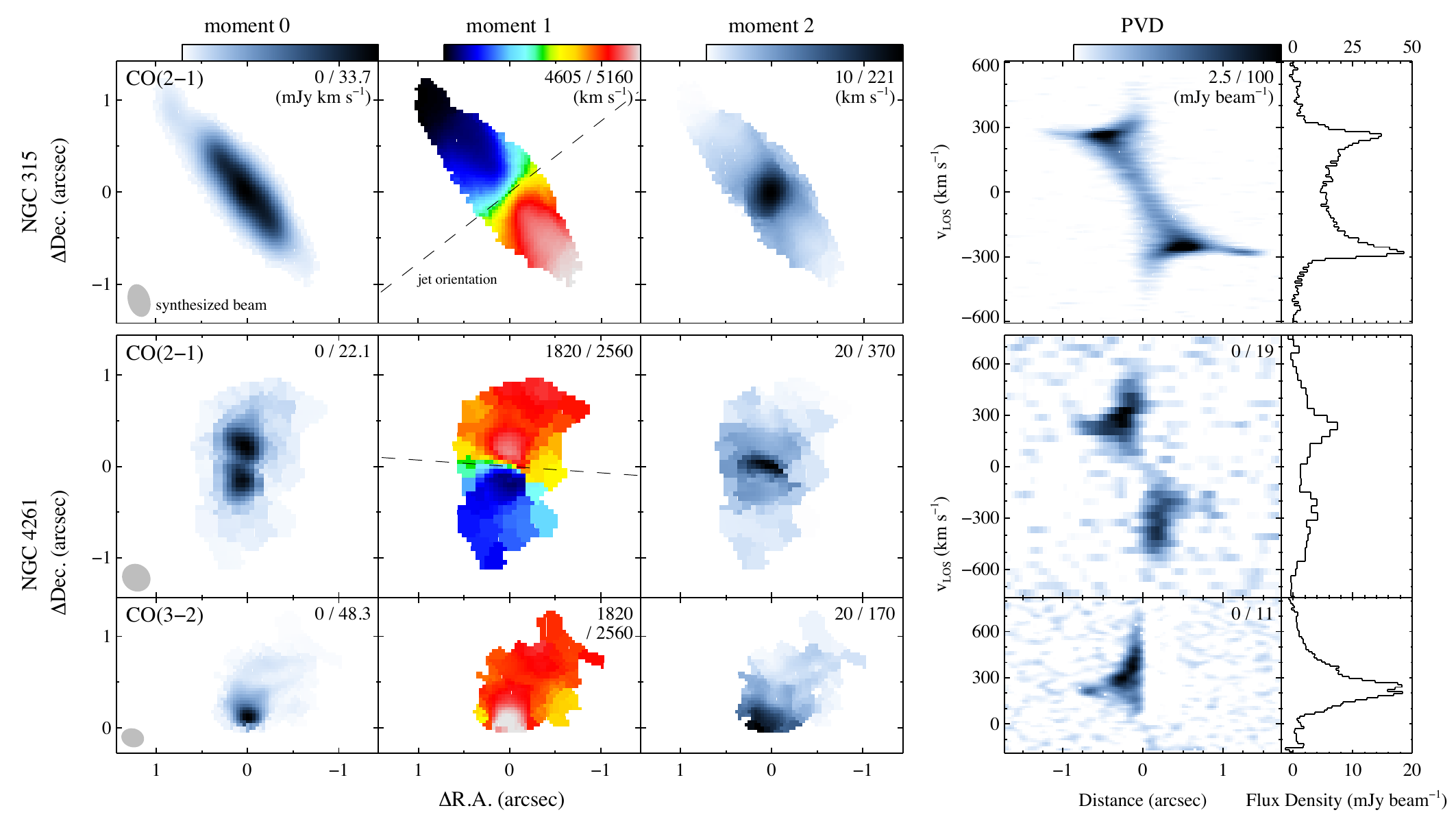}
\begin{singlespace}
  \caption{Moment maps constructed from NGC 315 \cotwo\ (\textit{top panels}) and NGC 4261 \cotwo\ and \cothree\ (\textit{middle/bottom panels}) ALMA data cubes. The moment maps reveal regular, albeit slightly warped, disk rotation. In the outer disk regions, spectra were binned together to achieve higher S/N. The moment values are linearly mapped to colors, shown by the color bar to the top of the maps, and the ranges in each panel indicate the minimum and maximum values of each color bar. The orientation of the radio jets (\textit{dashed lines}) are roughly perpendicular to the lines of nodes. Major-axis position-velocity diagrams (PVDs; \textit{right panels}) are displayed with their \vlos\ relative to \vsys\ from Table~\ref{tbl:tbl3}, with color bar ranges in each panel. Velocity profiles (\textit{far right panels}) for each cube were constructed by integrating flux densities within mask regions that follow the emission observed in each channel.\label{fig:fig6}}
\end{singlespace}
\end{center}
\end{figure*}

\subsection{Emission Line Properties\label{sec:almaproperties}}

The CO emission in the ALMA spectral cubes traces molecular gas in regular rotation about their respective galaxy centers. This CO emission originates from the same region as the arcsecond-scale dust disks detected in the \hst\ color maps, as shown in Figure~\ref{fig:fig5}. We constructed maps of the first three line moments, all of which are presented in Figure~\ref{fig:fig6}, after applying masks to exclude pixels that did not contain line emission. To ensure smooth first and second moment maps, especially in the outer disk regions, spectra were coadded in spatial bins to achieve higher S/N using a Voronoi tessellation technique \citep{cappellari03}.

We created CO position-velocity diagrams (PVDs) by extracting data from the NGC 315 cube along $\mathrm{PA}=39\degr$ and from the NGC 4261 cubes along $\mathrm{PA}=-5\degr$ using spatial extraction widths equal to the averages of their synthesized beam FWHMs. The CO emission reveals quasi-Keplerian rotation due to a central BH and an extended mass distribution. The gas velocities rise at projected distances of $\sim$15$-$30 pc from the galaxy centers, where the BHs dominate the gravitational potentials. The innermost \cotwo\ emission in NGC 315 has a maximum $\vlos\sim500$ \kms\ relative to \vsys, corresponding to $\mbh \sim 2.1\times 10^9$ \msun\ after assuming a disk $i\approx75\degr$. For NGC 4261, the innermost \cotwo\ and \cothree\ emission reaches $|\vlos-\vsys|\sim590$ and 660 \kms, respectively, suggesting $\mbh\sim(1.5-1.6)\times 10^9$ \msun\ for $i\approx65\degr$. The locii of high-velocity CO emission in the NGC 315 and NGC 4261 data sets remain more blended with minor-axis emission (with $\vlos\sim\vsys$) than the spectacularly well-resolved cases of NGC 3258 \citepalias{boizelle19} and NGC 383 \citep{north19}. The greater degree of blending is driven by both higher disk inclination angles and the factor of $\sim3$ larger $\theta_\mathrm{FWHM}$ of our Cycle 5 and 6 observations compared to those papers' data.

All three zeroth moment maps in Figure~\ref{fig:fig6} show apparently smooth, centrally concentrated CO emission, although the corresponding physical resolutions of 30$-$100 pc preclude any conclusions about CO substructure on cloud scales of a few to $\sim$30 pc \citep[e.g.,][]{utomo15,faesi18}. We measured a total \cotwo\ flux of $12.37\pm0.06(\mathrm{stat})\pm1.24(\mathrm{sys})$ Jy \kms\ for NGC 315 (quoting statistical and systemic uncertainties, respectively) and $3.06\pm0.15(\mathrm{stat})\pm0.31(\mathrm{sys})$ Jy \kms\ for NGC 4261. Due to the incomplete Band 7 spectral coverage, we cannot confidently estimate the total \cothree\ flux of the NGC 4261 disk. After degrading the \cothree\ cube angular resolution to match that of the NGC 4261 \cotwo\ data, we determined the CO line luminosities in temperatures units \citep[$L^\prime{}_\mathrm{CO}$;][]{carilli13} over the data sets' common velocity range and measured a ratio $R_{32}=L^\prime{}_\mathrm{CO32}/L^\prime{}_\mathrm{CO21}$ of $\sim$0.3$-$0.4 in the outer disk that rises to $\sim$0.6 near the nucleus. This rise in $R_{32}$ hints at a modest central increase in gas temperature and/or density.

\begin{deluxetable*}{ccccc|cccccc}
\tabletypesize{\scriptsize}
\tablecaption{Dynamical Model Properties\label{tbl:tbl2}}
\tablewidth{0pt}
\tablehead{\vspace{-1.5mm}\\
\multicolumn{5}{c}{NGC 315} & \multicolumn{5}{c}{NGC 4261}
}
\startdata
\vspace{-0.5mm}\multirow{2}{*}{Model} & \multirow{2}{*}{Mass Model} & $A_J$ & Disk Inclination & \multirow{2}{*}{$\sigmaturb(r)$} & \multirow{2}{*}{Model} & CO & \multirow{2}{*}{Mass Model} & $A_H$ & Disk Inclination & \multirow{2}{*}{$\sigmaturb(r)$} \\
 &  & (mag) & (deg.) & & & Transition &  & (mag) & (deg.) & \\\hline
A & MGE; free $\upsj$ & 0 & Free $i$ & Uniform & C1 & 2$-$1 & MGE; free \upsh & 0 & Free $i$ & Uniform \\
B1 & MGE; free $\upsj$ & 0 & Free $i$ & Gaussian & C2 & 2$-$1 & MGE; free \upsh & 0.4 & Free $i$ & Uniform \\
B2 & MGE; free $\upsj$ & 0.75 & Free $i$ & Gaussian & D & 2$-$1 & MGE; free \upsh & 0 & $i=60.8$ & Exponential \\
B3 & MGE; free $\upsj$ & 1.50 & Free $i$ & Gaussian & E & 3$-$2 & MGE; free \upsh & 0 & $i=60.8$ & Uniform
\enddata
  \tablecomments{Properties of the NGC 315 \cotwo\ and NGC 4261 \cotwo\ and \cothree\ gas-dynamical models, all of which treat the CO emission as arising from a thin, flat disk. Each galaxy's extended mass distribution was determined by deprojecting a $J$ or $H$-band MGE model from either Table~\ref{tbl:tbl1} or \ref{tbl:tbla}. The listed $A_J$ or $A_H$ refers to the assumed extinction correction to the portion of the stellar surface brightness originating behind the inclined dust disk. For NGC 4261, we examined two models with a molecular gas disk inclination angle fixed to the best-fit model C value to avoid implausibly low $i$: model D is identical to model C1 except for the different form used for \sigmaturb\ and model E is identical to model C1 except for its application to the \cothree\ data cube. While optimizing model E, we fixed $\vsys=2207.0$ \kms, as the \cothree\ data cube covers only the receding side of the disk.}
\end{deluxetable*}

We estimated H$_2$ gas masses assuming a line luminosity ratio $R_{21}=L^\prime{}_\mathrm{CO21}/L^\prime{}_\mathrm{CO10}=0.7$ and adopting a CO(1$-$0)-to-H$_2$ conversion factor $\alpha_\mathrm{CO}=3.1$ \msun\ pc\pertwo\ (K \kms)\per\ from a sample of nearby, late-type galaxies \citep{sandstrom13}. We then computed total gas masses ($M_\mathrm{gas}$) by correcting the H$_2$ mass measurements for the helium mass fraction $f_\mathrm{He}=0.36$. For NGC 315, the implied disk mass of $(M_\mathrm{gas}/10^8\,\msun)=2.39\pm0.01(\mathrm{stat})\pm0.24(\mathrm{sys})$ is consistent with previous upper limits \citep{davis19} and a tentative detection based on single-dish observations ($\log_{10}[M_\mathrm{gas}/\msun]\sim7.9$ after correcting for our assumed distance, $\alpha_\mathrm{CO}$, and $f_\mathrm{He}$; \citealp{ocana10}). In the same manner for NGC 4261, we estimated a total gas mass of $(M_\mathrm{gas}/10^7\,\msun)=1.12\pm0.05(\mathrm{stat})\pm0.11(\mathrm{sys})$ that is well below the detection threshold of previous surveys ($\log_{10}[M_\mathrm{gas}/\msun]\sim7.5-7.8$ after correcting for our assumed $\alpha_\mathrm{CO}$ and $f_\mathrm{He}$; \citealp{combes07,ocana10,young11}), and is therefore consistent with earlier non-detections. The peak gas mass surface densities ($\Sigma_\mathrm{gas}$) for the disks in NGC 315 and NGC 4261 are $3.1\times10^3$ \msun\ pc\pertwo\ and $1.5\times10^3$ \msun\ pc\pertwo, respectively, after deprojection. The $M_\mathrm{gas}$ and peak $\Sigma_\mathrm{gas}$ values are similar to those from other ETGs hosting few $\times100$-pc wide dust disks observed at similar physical resolutions (\citetalias{boizelle17}; \citealp{ruffa19a}).

To quantify potential deviations from purely circular, thin-disk rotation, we applied the \texttt{kinemetry} framework \citep{krajnovic11} to the CO first moment maps and traced the PA of the line-of-nodes ($\Gamma_\mathrm{LON}$) as a function of radius. From the NGC 315 disk edge to the center, the gas kinematic PA increases by only $\Delta\Gamma_\mathrm{LON}\sim7\degr$, consistent with measurements from other settled gas disks \citepalias[][\citealp{smith19}]{boizelle17}. We also calculated a global average $\overline{\Gamma}_\mathrm{LON}$ of $223\degr$ (for the receding side), which agrees with both the photometric and stellar kinematic PAs \citep{ene20} while being nearly perpendicular to the radio jet \citep{ensslin2001,lister18}. From the NGC 4261 \cotwo\ first moment map, we estimated $\Gamma_\mathrm{LON}\sim-23\degr$ at the disk edge, which is consistent with the photometric PA \citep[see also][]{krajnovic11} but is misaligned with the stellar kinematic PA \citep[e.g.,][]{davies86,krajnovic11}. Near the disk center, $\Gamma_\mathrm{LON}\sim-10\degr$ and is roughly perpendicular to the radio jet \citep{jones97}. Since $\Delta\Gamma_\mathrm{LON}>10\degr$ over the CO disk, and there is an offset in the centroids of the dust disk and stellar bulge \citep[Figure~\ref{fig:fig2}; also noted by][]{jaffe96,ferrarese96}, the NGC 4261 disk may not yet be fully settled into an equilibrium configuration.

\section{Dynamical Modeling\label{sec:dynmod}}

We determined BH masses from the ALMA data following a flat-disk forward-modeling procedure that is briefly summarized here. We refer the reader to the discussion by \citet{barth16a,barth16b} and \citetalias{boizelle19} for more details. After fitting models A$-$E (outlined in Table~\ref{tbl:tbl2}) to the corresponding CO data cubes, we explored sources of systematic uncertainty in the models to determine final error budgets for the NGC 315 and NGC 4261 BH masses.

\subsection{Method\label{sec:dynmod_method}}

We used models of the CO gas rotation and intrinsic turbulent velocity dispersion (\sigmaturb) to populate a model cube at each spatial location with Gaussian emergent line profiles. The models were fit directly to the ALMA CO data cubes, thereby allowing better characterization of model goodness-of-fit than if the fits were made to only moment maps. Model parameters were optimized by \chisq\ minimization using a downhill simplex approach \citep{press92}.

We began by calculating the circular velocity in the galaxy midplane as a function of radius. The circular velocity (\vc) arises from the combined gravitational potential of both an extended stellar mass distribution and a BH. We determined the stellar contribution to \vc\ by deprojecting the 2D light distribution parameterized by an MGE and scaling by the mass-to-light ratio $\Upsilon$. Given the typically small contribution of the molecular gas to the central gravitational potential in ETGs \citep[e.g.,][]{barth16b,davis17b,davis18,smith19}, we did not include the mass of the gas disk in the primary models A$-$E. However, we did explore the impact on the NGC 315 BH mass of including $M_\mathrm{gas}$ in the calculation of the total gravitational potential.

By projecting \vc\ for a given $i$ and $\Gamma$, with $\Gamma$ defined as the angle east of north to the receding side of the disk, we determined \vlos. Modeling the data cubes also requires assumptions about the intrinsic line widths. Previous ionized gas-dynamical models often required a central rise in the intrinsic line widths \citep[of $\gtrsim$100 \kms;][]{verdoes00,barth01,walsh10} to better match the observed kinematics. Thus, in addition to a spatially uniform $\sigmaturb=\sigma_1$, we explored two functional forms for the intrinsic gas velocity dispersion as a function of physical radius $r$ --  an exponential profile with $\sigmaturb(r)=\sigma_0\exp[-r/\mu]+\sigma_1$, and a Gaussian profile with $\sigmaturb(r)=\sigma_0 \exp[-(r-r_0)^2/2\mu^2]+\sigma_1$. The observed line widths are also the product of rotational broadening that arises from intra-pixel velocity gradients. To account for this effect, we oversampled the data cube pixel scale by a factor $s$ such that each pixel is divided into an $s\times s$ grid of elements. Both the model \vlos\ and \sigmaturb\ maps were calculated on this oversampled grid.

The line profiles were weighted by an approximation for the intrinsic CO surface brightness, formed using the IRAF STSDAS Richardson-Lucy deconvolution task \texttt{lucy} \citep{richardson72,lucy74}. We input the Voronoi-binned zeroth moment map and applied ten iterations of deconvolution, using the ALMA synthesized beam as the kernel. Since the CO surface brightness is not known on sub-pixel scales, we assumed each oversampled line profile has the same integrated flux and that the combined $s\times s$ total is equal to the deconvolved value at the native pixel scale. When optimizing the gas-dynamical model, we scaled this approximate flux map by a factor $f_0$ to account for possible normalization mismatches between the data and model.

The models outlined in Table~\ref{tbl:tbl2} have between seven and twelve free parameters each, including \mbh, a stellar mass-to-light ratio that is either \upsj\ or \upsh, disk $i$ and $\Gamma$ angles, $\sigmaturb(r)$, the kinematic center ($x_\mathrm{c}$,$y_\mathrm{c}$), a recessional velocity \vsys, and the scaling factor $f_0$. Before comparing the data and model cubes, we downsampled each $s\times s$ grid to the native pixel binning and convolved model cube channels with the ALMA synthesized beam. Since the noise in adjacent spatial pixels remains correlated, we elected to compute the model goodness-of-fit after spatially block-averaging the data and model in $4\times 4$ pixel regions to create nearly beam-sized cells. From the final rebinned data cube, we measured the rms background in line-free areas in each channel and calculated the \chisq\ statistic in a fitting region that fully encompasses the observed CO emission.

\begin{deluxetable*}{cccccccccccccr}
\tabletypesize{\scriptsize}
\tablecaption{Dynamical Modeling Results for NGC 315 and NGC 4261\label{tbl:tbl3}}
\tablewidth{0pt}
\tablehead{
Model & \colhead{\mbh} & \colhead{$\Upsilon$} & \colhead{$i$} & \colhead{$\Gamma$} & \colhead{$\sigma_1$} & \colhead{$\sigma_0$} & \colhead{$r_0$} & \colhead{$\mu$} & \colhead{$x_\mathrm{c}$} & \colhead{$y_\mathrm{c}$} & \colhead{\vsys} & \colhead{$f_0$} & \colhead{$\chisqnu$} \\
 & ($10^9$ \msun) & (\msun/\lsun) & ($\degr$) & ($\degr$) & (\kms) & (\kms) & (pc) & (pc) & (\arcsec) & (\arcsec) & (\kms) &  &  
}
\startdata
\multicolumn{14}{c}{\textbf{NGC 315}}\\
A & 2.39 & 2.06 & 74.1 & 218.1 & 15.3 & \nodata & \nodata & \nodata & $-0.007$ & 0.006 & 4969.0 & 0.970 & 2.046 \\
B1 & 2.40 & 2.05 & 74.1 & 218.1 & 15.2 & 147.3 & 4.39 & 32.0 & $-0.007$ & 0.005 & 4969.0 & 0.970 & 2.036\\
\multirow{2}{*}{\textbf{B2}} & 2.08 & 1.87 & 74.2 & 218.3 & 15.3 & 119.9 & 8.47 & 21.3 & $-0.007$ & 0.005 & 4969.1 & 0.958 & 1.992\\
 & (0.01) & (0.01) & (0.1) & (0.1) & ($^{+0.3}_{-0.2}$) & ($^{+9.4}_{-3.3}$) & ($^{+6.67}_{-3.43}$) & ($^{+6.3}_{-9.1}$) & (0.001) & (0.001) & (0.2) & (0.003) & \\
B3 & 1.96 & 1.88 & 74.2 & 218.3 & 15.2 & 83.0 & $-$0.12 & 35.3 & $-0.007$ & 0.005 & 4969.0 & 0.962 & 2.004 \\\hline
\multicolumn{14}{c}{\textbf{NGC 4261}}\\
\multirow{2}{*}{\textbf{C1}} & 1.67 & 1.62 & 60.8 & $-20.5$ & 40.6 & \multirow{2}{*}{\nodata} & \multirow{2}{*}{\nodata} & \multirow{2}{*}{\nodata} & 0.069 & 0.021 & 2207.0 & 1.050 & 1.263 \\
 & (0.10) & ($^{+0.46}_{-0.40}$) & ($^{+2.8}_{-3.3}$) & ($^{+2.0}_{-1.9}$) & (6.1) &  &  &  & (0.007) & (0.008) & ($^{+5.6}_{-5.5}$) & (0.032) & \\
C2 & 1.68 & 1.77 & 60.6 & $-20.8$ & 40.1 & \nodata & \nodata & \nodata & 0.070 & 0.023 & 2205.6 & 1.049 & 1.263 \\
D & 1.55 & 1.90 & \phantom{*}60.8* & $-15.5$ & 0.0 & 162.4 & \nodata & 72.2 & 0.053 & 0.017 & 2218.9 & 1.118 & 1.199 \\
E & 1.47 & 2.18 & \phantom{*}60.8* & $-17.9$ & 35.6 & \nodata & \nodata & \nodata & 0.040 & 0.036 & \phantom{*}2207.0* & 1.102 & 1.491
\enddata
  \tablecomments{Best-fit parameter values obtained by fitting the models described in Table~\ref{tbl:tbl2} to the \cotwo\ and \cothree\ data cubes. Parameter values followed by ``*'' were held fixed. For NGC 315, $\Upsilon$ is the $J$-band stellar mass-to-light ratio; for NGC 4261, $\Upsilon$ refers to the $H$-band mass-to-light ratio. The major axis position angle $\Gamma$ is measured east of north to the receding side of the disk. The disk kinematic center ($x_\mathrm{c}$, $y_\mathrm{c}$) is given in terms of right ascension and declination offsets from the nuclear continuum source centroid at 0$^{\rm h}$57$^{\rm m}$48\fs883, $+$30\degr21\arcmin08\farcs81 for NGC 315 and 12$^{\rm h}$19$^{\rm m}$23\fs216, $+$05\degr49\arcmin29\farcs69 for NGC 4261 (J2000). In these models, the disk systemic velocity  \vsys\ is taken to be the recessional velocity $cz_\mathrm{obs}$ in the barycentric frame that is used to transform the models to observed frequency units. Statistical uncertainties for the model B2 and C1 parameters are given in parentheses and were determined using the 68\% confidence intervals from Monte Carlo resampling.}
\end{deluxetable*}

\subsection{NGC 315 Modeling Results\label{sec:dynmod_results_ngc315}}

We fit models to the NGC 315 \cotwo\ data cube over a region that is elliptical in each channel and extends across velocities of 4400$-$5420 \kms\ (roughly $|\vlos-\vsys|\lesssim 550$ \kms). The spatial fitting region has a semi-major axis of $\rfit=1\farcs60$, an axis ratio of $b/a=0.26$, and a major-axis PA of $41\degr$. After block-averaging data and model cubes, the fitting region contains 10404 data points. We began by optimizing model A, which assumes a uniform \sigmaturb\ and employs the luminous mass model that was constructed after masking the most dust-obscured regions. The model A best-fit parameters are $\mbh=2.39\times 10^9$ \msun, $\upsj=2.06$ \msun/\lsun, and $\sigma_1=15.3$ \kms\ (see Table~\ref{tbl:tbl3} for the complete results). The total $\chisq=21272.0$ and the number of degrees of freedom ($\ndof$) is 10395, which results in $\chisqnu=\chisq/\ndof=2.046$.

\begin{figure*}[ht]
\begin{center}
\includegraphics[trim=0mm 0mm 0mm 0mm, clip, width=0.98\textwidth]{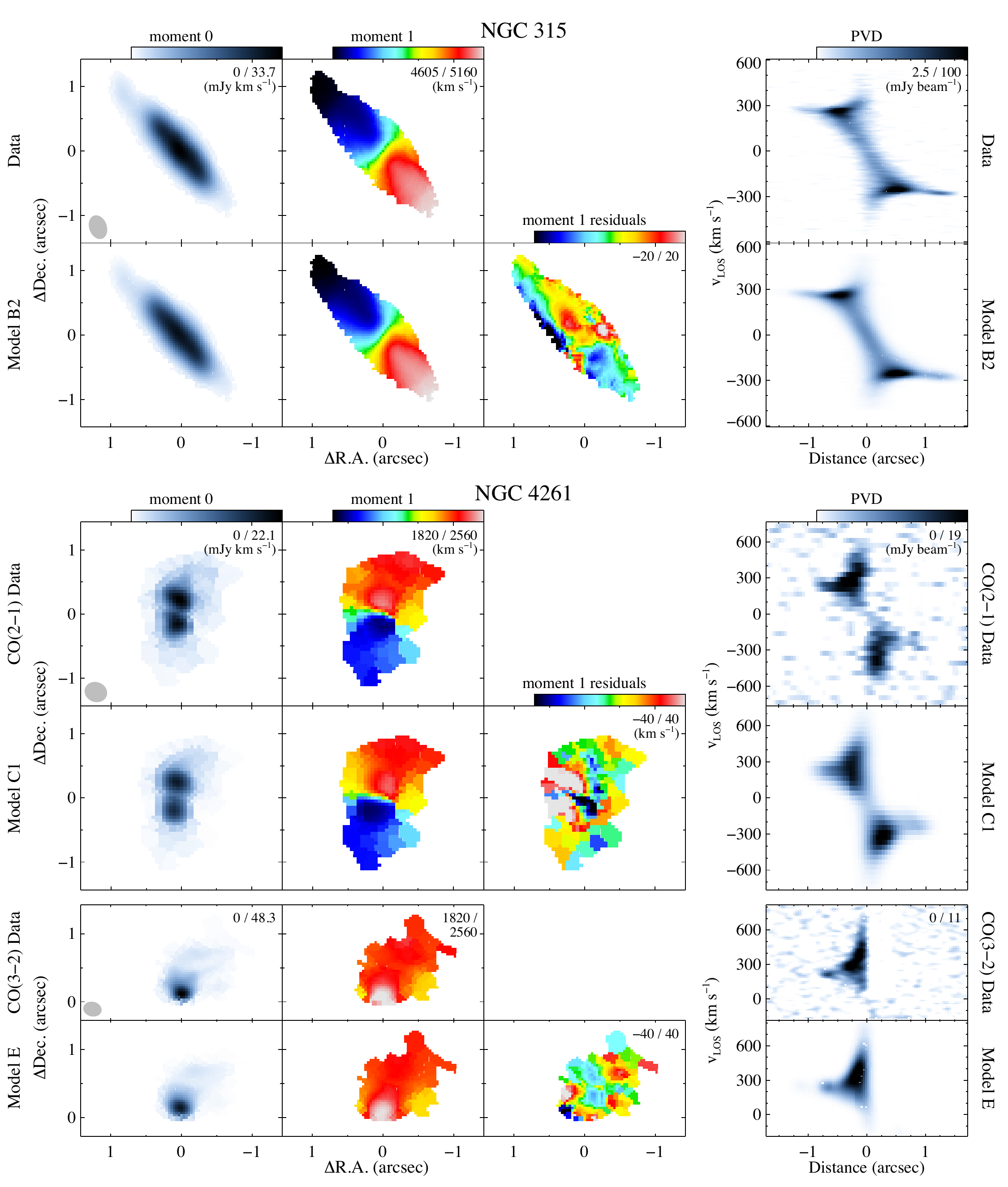}
\begin{singlespace}
  \caption{Comparison between the luminosity-weighted NGC 315 \cotwo\ (\textit{top panels}) and NGC 4261 \cotwo\ and \cothree\ (\textit{middle/bottom panels}) zeroth and first moment maps and PVDs with those derived from best-fitting models B2, C1, and E. With the noticeable exception of model C1 for NGC 4261, the first moment residual maps show generally small deviations ($\lesssim$10 \kms, or 4\%) between data and the best-fit models, with the largest discrepancies near the disk center. Likewise, the model PVDs generally agree with the data. However, model B2 for NGC 315 underrepresents the \cotwo\ emission that has $|\vlos-\vsys|\gtrsim350$ \kms, and model C1 for NGC 4261 cannot reproduce the sharp velocity upturn seen in the approaching side of the disk.\label{fig:fig7}}
\end{singlespace}
\end{center}
\end{figure*}

\begin{figure*}[ht]
\begin{center}
\includegraphics[trim=0mm 0mm 0mm 0mm, clip, width=\textwidth]{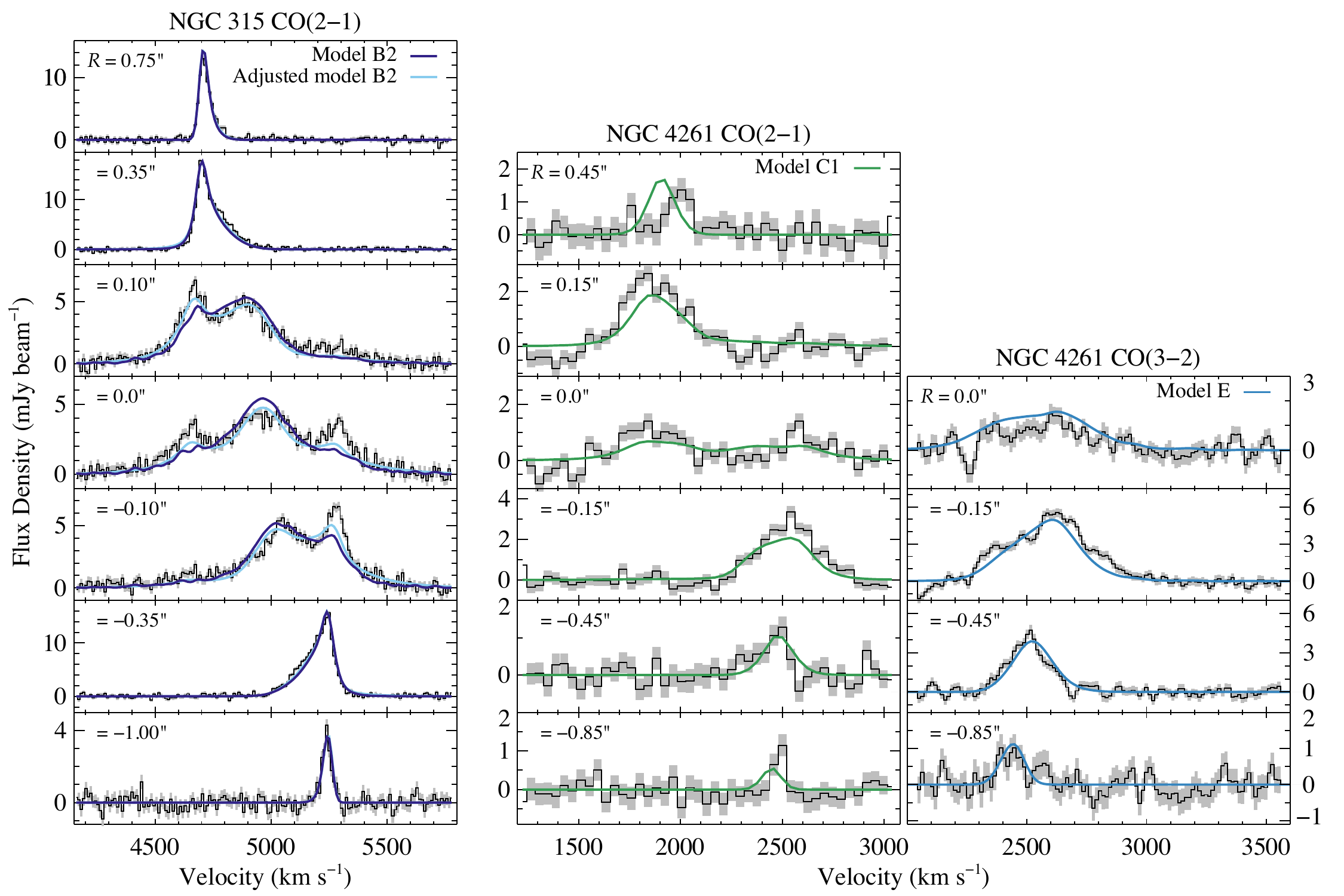}
  \caption{Comparison between Voronoi-binned line profiles extracted from both data and best-fit model cubes at select radii $R$ from the kinematic center along the disk major axis. Positive $R$ corresponds to the approaching side of the disk. Shaded regions indicate the frequency-dependent background rms. We adjusted the intrinsic flux map used in model B2 for NGC 315 and assigned larger CO flux within a central nuclear ring. The model line profiles are in better agreement with the data, and there is no change to \mbh. For NGC 4261, the discrepancies between the \cotwo\ data and model C1 at large radii are primarily driven by kinematic twists in the outer disk. In the central ($R=0\farcs0$) spectrum of the NGC 4261 \cothree\ data cube, there may be an absorption feature at $\vlos\sim2260$ \kms. Perhaps because of coarser channel binning, no absorption feature is apparent at the same location in the \cotwo\ data cube.\label{fig:fig8}}
\end{center}
\end{figure*}

For scenario B1, we used the same luminous mass model as in model A, while for B2 and B3 we adopted each of the extinction-corrected MGEs in turn. Models B1$-$B3 also included the more general Gaussian $\sigmaturb(r)$ prescription. Model B2 is the best match to the data, with $\chisq=20705.1$ over $\ndof=10392$, so we adopted B2 and its best-fit $\mbh=2.08\times10^9$ \msun\ as the fiducial model. Model B2 is not formally an acceptable fit with $\chisq_\nu=1.992$; however, the first moment map and PVD derived from the model B2 cube do match the observed CO kinematic behavior (Figure~\ref{fig:fig7}) apart from modest ($\pm20$ \kms) discrepancies near the kinematic center. For nearly the entire disk region (except for $R\lesssim0\farcs1$; Figure~\ref{fig:fig8}), the model line profiles closely follow the data. The best-fit $\upsj=1.87$ \msun/\lsun is below the expected 2.2<\upsj<2.8 \msun/\lsun\ for single stellar population (SSP) models that assume a \citet{salpeter55} initial mass function (IMF) with old ages ($\sim$10$-$14 Gyr) and solar metallicities, while the inferred 1.5<\upsj<1.9 \msun/\lsun\ from our dynamical model is above the expected $\upsj$ for a \citet{kroupa01} or \citet{chabrier03} IMF \citep{vazdekis10}.

Models B1 and B3 return $\mbh=(1.96-2.40)\times 10^9$ \msun\ and $\upsj=(1.88-2.05)$ \msun/\lsun, with $\chisq = 20829.9 - 21155.5$. All the NGC 315 models prefer a low $\sigma_1$ of $\sim$15.3 \kms, which is consistent with many other spatially resolved ALMA CO observations of ETGs (e.g., \citetalias{boizelle17}; \citealp{ruffa19b}). If we use a Gaussian function for \sigmaturb, the profile remains centrally peaked ($r_0\sim 0$ pc) and concentrated ($\mu\lesssim35$ pc), with $\sigma_0\sim 85-150$ \kms. Comparing models A and B1, the choice of a radially uniform versus a more flexible, centrally peaked \sigmaturb\ results in inconsequential changes to both \mbh\ and $\chisq$. For an even more edge-on disk in NGC 1332, \citet{barth16b} found a strong degeneracy between the $\sigma_0$ and \mbh\ parameters, concluding that beam-smearing effects were to blame. Subsequent gas-dynamical modeling of higher resolution data did not demonstrate that \sigmaturb\ needs to be centrally and broadly peaked to reproduce CO kinematics (\citealp{barth16a}; \citetalias{boizelle19}).

The $\sim$20\% difference in BH mass ($\Delta\mbh$) between models B1 and B3 is driven by a more than order of magnitude increase in the central stellar luminosity density. We treat $\Delta\mbh$ as representative of the systematic uncertainty introduced by circumnuclear dust. The $\Delta\mbh$ arising from dust corrections is roughly twice as large as the BH mass range derived from similar gas-dynamical models of NGC 3258 from ALMA Cycle 2 \cotwo\ observations \citepalias[with $A_H=0-1.50$ mag;][]{boizelle19}. The larger spread in BH masses for NGC 315 is the result of the disk's higher inclination angle and CO observations that do not extend as deeply within \rg.

\begin{figure*}[ht]
\begin{center}
\includegraphics[trim=0mm 0mm 0mm 0mm, clip, width=\textwidth]{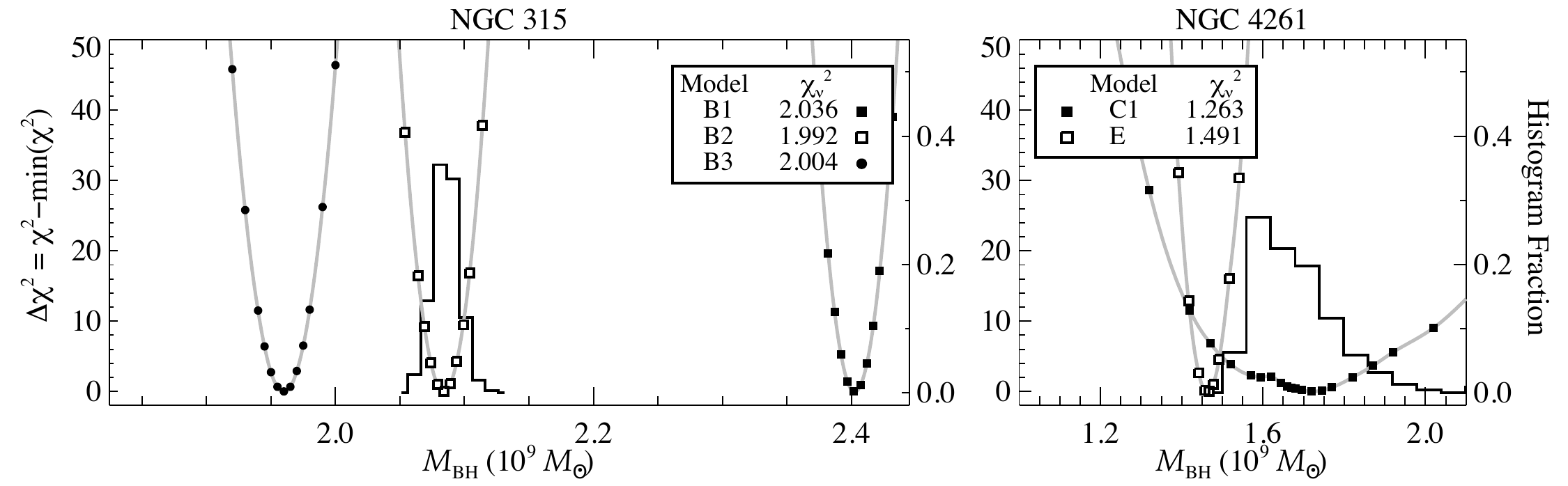}
\begin{singlespace}
  \caption{Results from \chisq\ minimization as a function of BH mass for the NGC 315 \cotwo\ and NGC 4261 \cotwo\ and \cothree\ dynamical models. Gray lines are spline interpolations to the $\Delta\chisq=\chisq-\min(\chisq)$ values. The histograms show the distributions of BH masses determined from Monte Carlo resampling of the B2 and C1 best-fit model cubes.\label{fig:fig9}}
\end{singlespace}
\end{center}
\end{figure*}

In Figure~\ref{fig:fig9}, we show $\Delta\chisq=\chisq-\min(\chisq)$ curves for models B1$-$B3. To construct the plots, we fixed the BH mass in each trial while allowing all other parameters to vary, and recorded the \chisq\ value from the best-fit model. The $1\sigma$ uncertainties derived from $\Delta\chisq<1$ regimes are $\sim$0.2\%, which are far smaller than the systematic uncertainties arising from plausible extinction corrections to the luminous mass model. Since block averaging does not fully eliminate noise correlations between neighboring pixels, we do not use $\Delta\chisq$ to determine the \mbh\ statistical uncertainty. Instead, we carried out 300 Monte Carlo realizations using the resampling procedure introduced in \citetalias{boizelle19}. At each iteration, line-free slices (where $|\vlos-\vsys|>560$ \kms) were drawn from the \cotwo\ data cube and randomly added to the best-fit model B2 cube before optimization of all model parameters. From this suite of Monte Carlo realizations, we estimated $1\sigma$ uncertainties for each parameter by taking the 15.9 and 84.1 percentiles of the respective distributions. The final statistical \mbh\ uncertainty is $\sim$10$^7$ \msun, or roughly 0.5\% of the best-fit BH mass. The remaining parameter statistical uncertainties for model B2 are listed in Table~\ref{tbl:tbl3}. We expect the statistical uncertainties on the model B2 parameters to be representative of those for the B1 and B3 models.

In addition to calculating the \mbh\ statistical uncertainty and exploring the effect of dust, we ran additional tests to measure the $\Delta\mbh$ that arises from other possible systematics. In each instance, we modified details of our model B2 to examine the impacts of various assumptions. These tests are described below.

\textit{Pixel oversampling:} Previous gas-dynamical modeling of resolved CO disks demonstrated little sensitivity to the choice of pixel oversampling (e.g., \citealp{barth16b}; \citetalias{boizelle19}). Nevertheless, we tested various factors from $s=1$ to $s=10$. The $\chisq$ decreased between $s=1$ and $s=2$, where $\chisqnu=1.904$, and then increased and plateaued at $\chisqnu\approx2.150$ for $s\geq 4$. For $s\geq4$, \mbh\ converged to the model B2 value in Table~\ref{tbl:tbl3}, with less than a 0.1\% scatter in the best-fit mass. Even without any pixel oversampling, the BH mass is only $\sim$1\% removed from the fiducial $s=4$ run.

\textit{Deconvolution:} At larger radii the best-fit models reproduce the observed emission line profiles very well. Closer to the nucleus, however, the non-unique deconvolution results in a nearly flat projected CO surface brightness, $\Sigma^\prime{}_\mathrm{CO}$, that does not give sufficient weight to high $|\vlos-\vsys|$ emission (see Figure~\ref{fig:fig8}). To explore the impact of an uncertain central CO distribution, we manually re-assigned deconvolved flux in the $s=4$ oversampled map, concentrating flux from the central 100 pc region into a tight elliptical annulus concentrated at radii of $\sim$30$-$40 pc. After rebinning to the native pixel scale and convolving with the beam, we find the re-assignment conserves flux and visually appears almost identical to the observed $\Sigma^\prime{}_\mathrm{CO}$. Using this adjusted flux map, the model fit returns a higher overall $\chisqnu = 2.046$ but better reproduces line profiles for $R<0\farcs1$ (see Figure~\ref{fig:fig8}). The resulting BH mass is lower by $\Delta\mbh=-7.4\times10^7$ \msun\ ($\sim$4\% of \mbh), consistent with studies that note the choice of intrinsic flux map impacts the quality of the fit but has a small effect on the inferred \mbh\ \citep[e.g.,][]{marconi06,walsh13}.

\textit{Gas mass:} Models A$-$B3 ignore the $M_\mathrm{gas}\approx 2.3\times 10^8$ \msun\ mass of the disk itself. We explored the impact on \mbh\ and \upsj\ by including the contribution to the circular velocity due to the gas mass ($v_\mathrm{c,gas}$) in addition to the stellar contribution derived from the MGE. We measured the \cotwo\ surface brightness from the zeroth moment map within elliptical annuli, determined the corresponding projected surface mass densities, and numerically integrated assuming a thin disk to calculate $v_\mathrm{c,gas}$. Throughout the disk, we found $v_\mathrm{c,gas}\lesssim$ 60 \kms. Including the gas mass in our dynamical model produced no noticeable change in \mbh, \upsj, or the other parameters.

\textit{Radial motion:} While the bright, unresolved mm-to-radio continuum source and radio jet suggests an actively accreting nucleus in NGC 315, the regular CO kinematics give no indication of significant non-circular gas motion near the BH. Regardless, we followed the method outlined in \citetalias{boizelle19} to estimate the potential impact of radial motion, and included a simple radial velocity term (\vrad) to represent either bulk inflow or outflow. The radial velocity term is a free parameter that is projected along the line of sight and added directly to the \vlos\ map. Although not fully self-consistent, this toy model provides an estimate of the amount of radial flow allowed by the data. After optimizing the model, we found a best-fit inflow speed of $\sim$11.4 \kms\ but a nearly identical inferred BH mass, within $\sim$0.3\% of model B2, and a somewhat improved $\chisqnu=1.959$. Radial flows produce kinematic twists in otherwise regular velocity fields. A similar twist can also be caused by a warped disk, and so the minor preference for radial inflow may instead be due to a slight warp in the gas disk.

\textit{Final error budget:} As expected, the systematic uncertainties (sys) in the NGC 315 BH mass measurement are dominated by the uncertainty due to the dust correction. To this term, we added the remaining $\Delta\mbh$ from the other systematic effects we explored in quadrature. Together with the statistical (stat) BH mass uncertainty estimated from Monte Carlo realizations, our final BH mass measurement with $1\sigma$ uncertainty ranges is $(\mbh/10^9\,\msun)=2.08\pm0.01\,(\mathrm{stat})\,^{+0.32}_{-0.14}\,(\mathrm{sys})$. Our quoted systematic error budget for NGC 315 is provisional given that our models do not provide a formally acceptable fit to the data. We have chosen not to examine \textit{ad hoc} procedures to address this issue (e.g., degrading the data quality by inflating the error bars to achieve a reduced \chisq of 1 and therefore a formally acceptable fit, or excluding portions of the data where the model fits are poor), and these issues will be mitigated by future (approved Cycle 7) higher resolution data that will allow for a less model-dependent approach.

\subsection{NGC 4261 Modeling Results\label{sec:dynmod_results_ngc4261}}

We optimized the thin-disk model C1 to the NGC 4261 \cotwo\ cube, with spatially uniform \sigmaturb\ and an MGE constructed from the dust-masked $H$-band image. This fit was made to an elliptical spatial region that is uniform across channels with velocities in the range $1585-2805$ \kms\ (roughly $|\vlos-\vsys|\lesssim 600$ \kms) with $\rfit=1\farcs00$, an axis ratio of $b/a=0.42$, and a major-axis $\mathrm{PA}$ of $-20\degr$. After $4\times 4$ block-averaging the data and model cubes, the fitting region contains 1178 data points. This model, with a best-fit BH mass of $1.67\times10^9$ \msun, achieved $\chisq=1476.7$ over $\ndof=1169$ for $\chisqnu=1.263$. The best-fit $\upsh=1.62$ \msun/\lsun\ is consistent with the expected $H$-band stellar mass-to-light ratios for old, solar metallicity SSPs ranging between 1.02$-$1.84 \msun/\lsun\ depending on the IMF \citep{vazdekis10}.

\begin{figure}
\begin{center}
\includegraphics[trim=0mm 0mm 0mm 0mm, clip, width=\columnwidth]{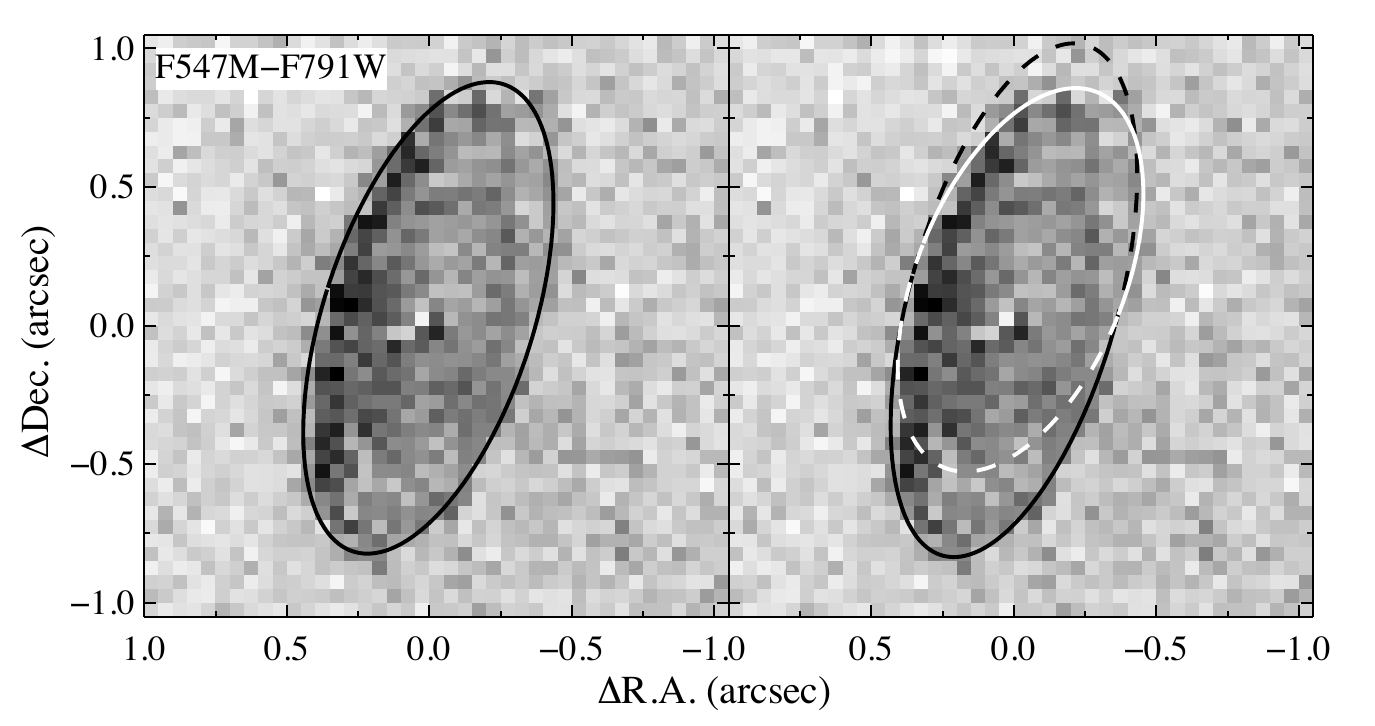}
\begin{singlespace}
  \caption{Central $2\arcsec\times 2\arcsec$ region of the NGC 4261 F547M$-$F791W color map used to estimate the disk inclination angle. Superposed on the image are ellipses fit to color contours that trace the entire outer disk (\textit{left panel}) and those fit to either the receding (\textit{white}) or the approaching (\textit{black}) side. The dashed lines show where the ellipses are extrapolated beyond the fitting regions.\label{fig:fig10}}
\end{singlespace}
\end{center}
\end{figure}

The best-fit $i=60.8\degr$ is slightly lower than the value estimated by \citet{ferrarese96}, who used the outer dust disk $b/a$ from the F547M$-$F791W color map to estimate $i\approx\cos\per (b/a)=64\degr$. They noted the disk has different oblateness on the approaching and receding sides, so we extended their analysis by also fitting additional ellipses to the color contours on each side of the disk separately (Figure~\ref{fig:fig10}). Our full-disk fit returns $b/a=0.42$, while $b/a=0.38$ and 0.49 for the receding and approaching sides of the disk, respectively. The corresponding disk inclination angle is $(65.3^{+2.1}_{-4.7})\degr$, and we estimated a major-axis PA for the disk of $(-17.2^{+2.1}_{-3.7})\degr$. Our model C1 $i$ and $\Gamma$ appear consistent with the dust disk morphology and orientation.

The best-fit model first moment map qualitatively agrees with the observed CO kinematic behavior at most locations, although discrepancies exceeding 40 \kms\ arise both near the nucleus and at the disk edge, as our thin-disk formalism does not allow for a changing disk PA with radius (see Figures~\ref{fig:fig5} and \ref{fig:fig8}). In \citetalias{boizelle19}, we examined a gas disk with a similar $\Delta\Gamma_\mathrm{LON}\sim20\degr$ and demonstrated that accounting for a slight warp in the disk may introduce a shift in BH mass at the few-percent level. The model C1 PVD does not fully capture the sharp central rise in \cotwo\ emission-line speeds, most noticeably on the approaching side of the disk. One plausible explanation is a rapid increase in disk inclination angle in the central $\sim$40 pc. A second possibility is a sudden rise in intrinsic line widths from $\sim$20 to over 100 \kms\ at the same radius. The S/N of the NGC 4261 \cotwo\ data does not permit exploration of a more general disk structure, but we do test a more flexible \sigmaturb\ function in model D.

The model C1 $\Delta\chisq$ as a function of BH mass (Figure~\ref{fig:fig9}) suggests a much larger statistical uncertainty compared to the NGC 315 results. As we increased the fixed $\mbh$ above $\sim$1.$8\times10^9$ \msun, the best-fit $i$ trended downwards toward $\sim$55\degr, resulting in a noticeably asymmetric $\Delta\chisq$ curve. Following the NGC 315 approach, we estimated the NGC 4261 statistical uncertainty by carrying out 300 Monte Carlo realizations, randomly adding line-free channels of the \cotwo\ data cube to the best-fit model C1 cube. We determined the $1\sigma$ uncertainty for each parameter from the respective distributions (Table~\ref{tbl:tbl3}); the \mbh\ statistical uncertainty is $1.0\times10^8$ \msun, or 6\% of the model C1 BH mass. Notably, the statistical uncertainties for \upsh\ and $\sigma_1$ are much larger, exceeding 25\% and 15\% of the respective best-fit values. Also, the statistical uncertainty on $i$ ($\pm3\degr$) is larger than typically seen with ALMA CO dynamical models, which is likely the result of our not having high S/N CO emission over the full spatial extent of the disk.

Next, we employed the extinction-corrected ($A_H=0.4$ mag) MGE in model C2. As demonstrated in \citetalias{boizelle19} and above for NGC 315, the uncertainty in the dust attenuation can be the dominant systematic of the BH mass error budget. However, the model C2 $\chisq=1476.3$, which is essentially identical to that of model C1. The best-fit C2 model has $\mbh=1.68\times10^9$ \msun\ and $\upsh=1.77$ \msun/\lsun, indicating that a ``dust-corrected'' stellar mass model has negligible impact on the best-fit parameters. The stark difference in $\Delta\mbh$ for NGC 315 and NGC 4261 suggests that a larger portion of the NGC 4261 gas disk is within \rg, and therefore the central CO kinematics are much less sensitive to the extended mass distribution.

We discussed above how the best-fit C1 model qualitatively agrees with the observed first moment values at most locations, with the largest discrepancies occurring near the nucleus and the disk edge. In addition to these discrepancies, the model C1 best-fit $\sigma_1$ of $40.6$ \kms\ results in model line profiles that are excessively broad near the disk edge. In model D, we relaxed the radially-uniform constraint on intrinsic line widths by adopting an exponential function for \sigmaturb. Early model D trials produced better fits with $\chisqnu<1.17$ and substantially higher (by at least 10\%) \mbh\ and \upsh. They also preferred disk inclination angles below 45\degr, which are highly inconsistent with the dust disk morphology and too low to allow for deprojection of the MGE. Closer inspection revealed a broad degeneracy between $i$ and \mbh. We thus fixed the model D $i$ to the best-fit value from model C1 before optimizing the remaining free parameters. However, we note that the bulk of the CO emission is concentrated well within the outer edge of the dust disk that is used to estimate $i$ (see Figure~\ref{fig:fig5}). Therefore, the dust morphology does not preclude a large decrease in $i$ towards the disk center. For model D, the best-fit BH mass decreases to $1.55\times10^9$ \msun\ while \upsh\ increases to 1.90 \msun/\lsun. The \sigmaturb\ profile has $\mu = 72.2$ pc and line-width amplitudes of $\sigma_0 = 162.4$ \kms\ and $\sigma_1 = 0.0$ \kms. At the disk edge, $\sigmaturb\sim20$ \kms, which is in closer agreement with the observed line widths than the uniform $\sigma_1$ value found for models C1$-$C2. Even though model D is a better fit to the \cotwo\ data with $\chisq=1400.4$, because $i$ is not a free parameter we elected to retain model C1 as our fiducial gas-dynamical model for NGC 4261. As a final note, the model D best-fit \vsys\ is 12 \kms\ larger than the model C1 value. The shift is likely due to the combination of the coarse channel spacing, the different line-width prescription, and low S/N in the outer disk.

We investigated the stability of our \cotwo\ gas-dynamical modeling results by fitting the \cothree\ data cube with thin-disk model E that is analogous to model C1. Since the frequency coverage of the \cothree\ data excludes most of the approaching side of the disk, we cannot constrain the systemic velocity, and we fixed this parameter to the best-fit \vsys\ from model C1. The spatial ellipse of the fitting region remains the same as for the \cotwo\ models, but we adopt a smaller velocity range of 2035$-$2920 \kms\ to calculate \chisq. After $4\times 4$ block-averaging the data and model cubes, the fitting region contains 5280 data points. Like model D, early runs of model E showed a preference for very low $i$ values, so we also fixed the disk inclination to the model C1 value. Optimizing all remaining free parameters gives $\mbh=1.47\times10^9$ \msun, $\upsh=2.18$ \msun/\lsun, and $\sigma_1=35.6$ \kms, with a total $\chisq=7863.0$ over $\ndof=5273$ for $\chisqnu=1.491$. In Figure~\ref{fig:fig7}, we compare \cothree\ moment maps and the PVD constructed from the best-fit model E with those drawn from the data cube. Line profiles extracted from the best-fit cube show good agreement with the data (Figure~\ref{fig:fig8}). Because model E adopts a fixed $i$, the curve of $\Delta\chisq$ as a function of \mbh\ is narrow and symmetrical (Figure~\ref{fig:fig9}).

The range of BH mass measurements found from models C1$-$E of $(1.47-1.68)\times10^9$ \msun\ reflect a few fundamental differences in model construction. The following paragraphs describe additional tests performed to probe other sources of systematic uncertainty when modeling the NGC 4261 \cotwo\ data cube. In each instance below, we modified aspects of our model C1 to estimate the impact of each systematic effect on the best-fit \mbh.

\textit{Pixel oversampling:} As was the case for NGC 315, our NGC 4261 dynamical models converge to essentially the same BH masses when we adopt an oversampling of $s\geq4$. For $s=1$, the inferred BH mass is $\Delta\mbh=-2.0\times10^8$ \msun\ compared to the fiducial value with a lower corresponding $\chisqnu=1.241$. With increasing oversampling, \chisqnu\ increases until reaching a plateau at $s\geq6$ with $\chisqnu\approx 1.285$. For $s\geq4$, we found a scatter of $2.1\times10^7$ \msun\ ($\sim$1.5\%) in BH mass. 

\textit{Block averaging:} The decision to $4\times4$ block-average data and model cubes before calculating \chisq\ mitigates noise correlation between spatial pixels at the expense of spatial sampling. For the \cotwo\ data, our approach leaves fewer than ten points per slice for $R<0\farcs5$. However, within $0\farcs5$ the highest S/N CO emission and the most distinct Keplerian rotation signature is found. When the block-averaging step is instead skipped, fits to the native data cube converged on a similar \mbh\ of $1.74\times10^9$ \msun, corresponding to $\Delta\mbh=7.0\times10^7$ \msun\ or a 4\% increase from the best-fit model C1 value. Our choice of block averaging size thus did not significantly bias the BH mass measurement.

\textit{Disk inclination:} The best-fit $i=60.8\degr$ from model C1 is at the lower end of the confidence interval 60.6\degr$-$67.4\degr\ derived from the morphology of the dust disk. Given the large estimate for the statistical uncertainty in the inclination angle, we ran a test with a fixed $i=67.4\degr$ that is at the high end of the likely inclination angles. In this case, the BH mass was higher by $3.7\times10^8$ \msun\ (or $\sim$22\%).

\textit{Radial motion:} The slightly disturbed CO kinematics and irregular dust disk morphology may signal non-negligible radial gas motion in the NGC 4261 disk. Introducing a bulk radial flow term \vrad\ as was done for NGC 315, we found a slight preference for an outflow with $\vrad\sim$25 \kms, which is only $\sim$10\% of the observed CO \vlos\ at the disk edge. Compared to model C1, the \chisq\ and the best-fit \mbh\ increase by 6.7 and $5.5\times10^6$ \msun, respectively. We note that the preference for $\vrad\neq0$ \kms\ is in part driven by tension between our axisymmetric thin-disk model and the observed CO kinematic twists.

\textit{Fitting region:} By applying \texttt{kinemetry} to the \cotwo\ first moment map, we found that $\Delta\Gamma_\mathrm{LON}=13\degr$ from the disk center to the edge. To determine the potential impact on \mbh\ of the moderate kinematic twist, we ran a test that restricted \rfit\ to 0\farcs55. Within this fitting region, the \texttt{kinemetry} results suggest that the gas is in simple, thin-disk rotation with $\Delta\Gamma_\mathrm{LON}<3\degr$. Data within the smaller \rfit\ are insensitive to the extended stellar mass contributions, so we fixed \upsh\ to the model C1 value. With only ten data points per channel and no high S/N CO detections that lie off the major axis, the inclination angle also needed to be fixed to 60.8\degr. When using the more central fitting region, we find that the best-fit BH mass increases by $\Delta\mbh=8.0\times10^7$ \msun\ or $\sim$5\%.

\textit{Final error budget:} The largest positive and negative systematics for the NGC 4261 ALMA CO gas-dynamical models are the 22\% increase in BH mass when adopting the upper-bound inclination angle and the 12\% decrease when changing from \cotwo\ to \cothree\ kinematic modeling. While not necessarily independent, we included the remaining systematic terms using a quadrature sum. Our final BH mass with $1\sigma$ statistical uncertainties is $\log_{10}(\mbh/10^9\,\msun)=1.67\pm0.10\mathrm{(stat)}^{+0.39}_{-0.24}\mathrm{(sys)}$.

\section{Discussion\label{sec:discussion}}

From ALMA observations of rotating circumnuclear molecular gas disks in the active galaxies NGC 315 and NGC 4261, we infer the presence of BHs with respective masses of $2.08\times10^9$ and $1.67\times10^9$ \msun. Our work provides the first dynamical measurement of the BH in NGC 315 and significantly improves upon a prior ionized gas-dynamical \mbh\ determination for NGC 4261 \citep{ferrarese96}. Both galaxies were observed as part of our ongoing program to take an accurate census of BHs in massive ETGs by exploiting ALMA's unique capabilities. In the cases of NGC 315 and NGC 4261, we obtain $\sim$10$-$25\% precision on \mbh, with the final error budgets being dominated by modeling systematics. Below we compare our determinations to other BH mass estimates, and discuss how well the ALMA observations resolve the NGC 315 and NGC 4261 BH spheres of influence, possible future improvements to the \mbh\ measurements, and prospects for inferring BH masses in active galaxies with ALMA.

\subsection{BH Mass}

\begin{figure}[ht]
\begin{center}
\includegraphics[trim=0mm 0mm 0mm 0mm, clip, width=\columnwidth]{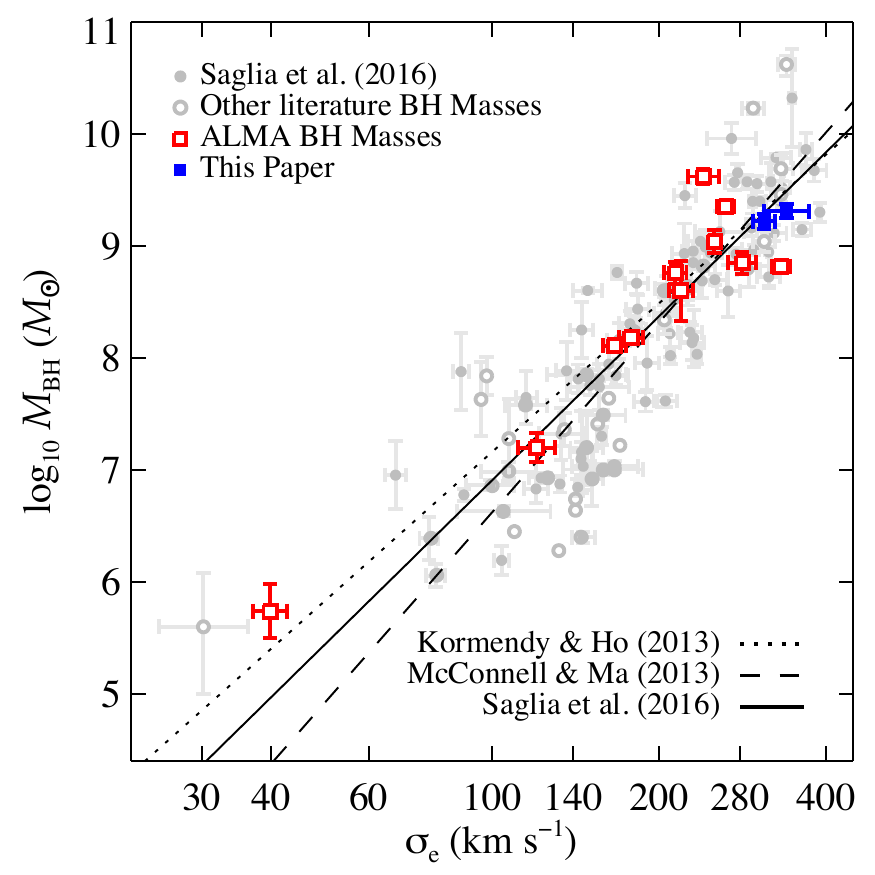}
\begin{singlespace}
  \caption{Comparison between dynamically measured BH masses and host galaxy $\sigma_\mathrm{e}$ \citep[\textit{filled gray circles}:][]{saglia16}, along with other literature values \citep[\textit{open gray circles}:][]{denbrok15,greene16,thomas16,walsh16,walsh17,erwin18,krajnovic18,mehrgan19,liepold20} and those measured using ALMA observations \citep[\textit{colored squares}:][]{barth16a,onishi17,davis17b,davis18,north19,smith19,nagai19,boizelle19,ruffa19b,davis20,nguyen20}. When not listed in the papers, we adopted $\sigma_\mathrm{e}$ values and/or uncertainties when available from other sources \citep{barth02,filippenko03,brough07,cappellari13a,kormendy13,ma14,saglia16,vandenbosch16b,veale17}. The NGC 315 and NGC 4261 BH masses presented here are consistent with $\mbh-\sigma_\mathrm{e}$ relations from \citet{kormendy13}, \citet{mcconnell13a}, and \citet{saglia16}.\label{fig:fig11}}
\end{singlespace}
\end{center}
\end{figure}

Using the stellar velocity dispersion measured within the galaxy effective radius ($\sigma_\mathrm{e}$) for NGC 315 of 341 \kms\ \citep{veale17}, the $\mbh-\sigma_\star$ relation \citep{kormendy13,mcconnell13a,saglia16} predicts BH masses in the range of $(3.2-3.9)\times 10^9$ \msun\ (Figure~\ref{fig:fig11}), where we have adopted the fit to elliptical galaxies and classical bulges from \cite{kormendy13} and \cite{saglia16} and the fit to early-type galaxies from \cite{mcconnell13a}. Likewise, the expected \mbh\ for NGC 4261, with $\sigma_\mathrm{e} = 315$ \kms\ \citep{kormendy13}, is $(2.2-2.7)\times10^9$ \msun. Our dynamical \mbh\ measurements for both galaxies are below the mean values predicted from the $\mbh-\sigma_\star$ relation, but are within the intrinsic scatter of the relation. Even if we assume a smaller stellar velocity dispersion of $\sigma_\mathrm{e} = 265$ \kms\ for NGC 4261 \citep{cappellari13a}, our dynamical \mbh\ remains consistent with the $\mbh-\sigma_\star$ relation.

In order to compare to the $\mbh - L_\mathrm{bul}$ relation \citep{kormendy13}, we used the MGE models in Table~\ref{tbl:tbl1} and assumed colors of $J-K=0.9$ mag and $H-K=0.2$ mag for an old, solar metallicity SSP \citep{vazdekis10}. We found $L_{\mathrm{bul},K} = 8.9\times10^{11}$ \lsun\ for NGC 315 and $1.3\times10^{11}$ \lsun\ for NGC 4261, corresponding to predicted BH masses of $7.8\times10^9$ \msun\ and $0.72\times10^9$ \msun, respectively. If instead we used the total apparent $K$-band magnitude from the HyperLeda database \citep{makarov14} for NGC 315, then $L_{\mathrm{bul},K} = 6.8\times10^{11}$ \lsun, which translates to $\mbh = 5.7\times10^9$ \msun. The $L_{\mathrm{bul},K}$ estimated from our MGE model for NGC 4261 agrees with the HyperLeda value. Thus, our NGC 315 gas-dynamical \mbh\ is an outlier, lying below the lower envelope of BH masses populating the $\mbh-L_{\mathrm{bul}}$ relation, while our ALMA-based \mbh\ for NGC 4261 is consistent with the $\mbh-L_{\mathrm{bul}}$ correlation.

We also estimated the bulge mass by multiplying the $J$ and $H$-band total luminosities from the MGEs in Table~\ref{tbl:tbl1} by \upsj\ from model B2 for NGC 315, and by \upsh\ from model C1 for NGC 4261. This estimate resulted in a bulge mass of $1.2\times10^{12}$ \msun\ for NGC 315, a predicted $\mbh = (3.2-3.9)\times10^9$ \msun\ from \cite{mcconnell13a} and \cite{saglia16}, and $\mbh = 9.0\times10^9$ \msun\ from \cite{kormendy13}. For NGC 4261, we found a bulge mass of $3.8\times10^{11}$ \msun\ and an expected $\mbh = (1.2-2.3)\times10^9$ \msun\ \citep{kormendy13,mcconnell13a,saglia16}. Our NGC 315 BH mass is consistent within the intrinsic scatter of the $\mbh - M_\mathrm{bul}$ relation \citep{mcconnell13a,saglia16} but lies well below the predicted value using the $\mbh - M_\mathrm{bul}$ relation from \cite{kormendy13}, although the large uncertainty in the galaxy distance limits the  significance of the discrepancy. Our NGC 4261 BH mass is consistent with all of the $\mbh - M_\mathrm{bul}$ relations.

At present, there is not another dynamical mass measurement for the NGC 315 BH. However, \citet{beifiori09} estimated a rough BH mass upper limit by fitting a rotating disk model to the central [\ion{N}{2}] line width, which was measured from a single spectrum extracted from an \hst\ Space Telescope Imaging Spectrograph (STIS) observation. They neglected the stellar contribution to the gravitational potential and did not consider the effects of dynamically significant turbulent motion. Assuming an inclination angle of $81\degr$ and $33\degr$, they found \mbh\ upper limits of $(0.5-2.0)\times10^9$ \msun\ (scaled to our adopted distance). While the STIS data appear to support a conclusion of rotation \citep{noelstorr03}, the ionized gas kinematics are sufficiently chaotic to remove any tension with our ALMA-derived BH mass. Possible future stellar-dynamical modeling of this galaxy \citep{ma14,ene19} may enable a meaningful direct comparison between a precision ALMA CO BH mass and a stellar-dynamical BH mass measurement \citep[e.g.,][]{krajnovic09,rusli11,schulze11,barth16a,davis17b,smith19}.

In contrast to NGC 315, the mass of the central BH in NGC 4261 has been previously measured using ionized gas. \citet{ferrarese96} conducted one of the earliest gas-dynamical BH studies with \hst, and the measurement was used to establish the original $\mbh-\sigma_\star$ relation along with a small handful other targets \citep{gebhardt00,ferrarese00}. \citet{ferrarese96} derived gas kinematics from 13 nuclear spectra obtained with the \hst\ Faint Object Spectrograph (FOS), and modeled the radial velocities assuming a purely Keplerian potential. They found a BH mass of $(5.0 \pm 1.0)\times10^8$ \msun\ (scaled to our distance). Although our ALMA gas-dynamical \mbh\ is inconsistent with \citet{ferrarese96}, we employed a much more sophisticated and detailed dynamical model using methods that have been developed over the past two decades. In particular, the ionized gas-dynamical modeling did not account for PSF blurring, which is expected to yield an underestimate of the BH mass. Also, \citet{kormendy13} argue that some of the BH masses based on ionized gas, including NGC 4261, were likely underestimated because of non-gravitational gas perturbations. The ALMA data further provide substantial advantages over the FOS observations, including better spatial coverage of the gas disk and a focus on a well-defined, isolated CO emission line. We note that \citet{humphrey09} estimated a BH mass that was consistent with \citet{ferrarese96} by modeling the hot interstellar medium of NGC 4261, assuming the X-ray emitting gas is in hydrostatic equilibrium. However, tension with our \mbh\ suggests the hot plasma near the AGN is strongly affected by non-gravitational motions.

\subsection{Resolving the BH Sphere of Influence\label{sec:rgdiscussion}}

The precision of a BH mass measurement is in large part determined by how well the BH sphere of influence is resolved. In \citetalias{boizelle19}, we detected CO(2-1) down to radii of $\sim0.14\rg$ in NGC 3258. The ALMA data had a relative resolution of $\xi=2\rg/\theta_\mathrm{FWHM}\sim17$ and facilitated an \mbh\ measurement with percent-level precision. A relative resolution of $\xi\sim17$ is typical of very long-baseline interferometry observations of megamaser galaxies \citep[e.g.,][]{kuo11,zhao18}. Although a few similar cases exist \citep{smith19,north19,nagai19}, most published ALMA CO data sets of circumnuclear disks are obtained with $\xi\lesssim2$ \citep[e.g.,][]{onishi17,boizelle17,davis17b,davis18,ruffa19a,zabel19,nguyen20}. This limitation leads to larger uncertainty in the BH mass measurement, often driven by various systematic effects. In addition, spatial blurring in highly inclined disks entangles minor-axis ($\vlos\sim\vsys$) emission with the highest $|\vlos-\vsys|$ emission near the nucleus \citep[][]{barth16b}. In order to evaluate how well the data resolve \rg, one should also consider the relative resolution of \rg\ along the projected minor axis, or if $\xi\cos i \gtrsim 2$ \citep{barth16b}. To date, only three ALMA CO observations meet this criterion and detect emission deep within the BH-dominated region (\citealp{nagai19}; \citetalias{boizelle19}; \citealp{north19}).

For the NGC 315 \cotwo\ data, we directly calculated \rg\ for models A$-$B3 by finding the radius where the various luminous mass contributions to \vc\ equal that from the BH. We find that \rg\ ranges between 0\farcs67$-$0\farcs80 with CO emission detected down to $\sim$(0.11$-$0.14)\rg. This result corresponds to a high $\xi$ of $\sim$5, but only a marginally resolved $\xi\cos i$ of $\sim1.4$. For NGC 4261, models C1$-$E suggest $\rg=1\farcs24-1\farcs58$, which extends beyond the disk edge. We detected CO down to radii of $\sim$(0.06$-$0.08)\rg. With $\xi\sim9.6$ and $\xi\cos i$ of about 4.9, the current CO data fully resolve the BH sphere of influence.

The dominant systematic in the NGC 315 measurement is the uncertainty in the luminous mass model due to the presence of nuclear dust. Using different MGE models with various levels of extinction correction leads to a $\sim$6$-$15\% uncertainty in \mbh. This exercise highlights the need to carefully consider the effects of dust in situations where the ALMA observations only marginally resolve \rg. For NGC 315, \rg\ is well resolved along the major axis but not well resolved along the projected minor axis. We expect the systematic uncertainty due to dust to be even more severe for cases where \rg\ along the major axis is also not well resolved. Higher angular resolution CO imaging of NGC 315, approved in an ALMA Cycle 7 program, will fully isolate the locus of rapid gas rotation within \rg\ and allow for the extended stellar mass distribution to be constrained directly from the CO emission-line kinematics, as was done in \citetalias{boizelle19}. Such an approach will eliminate the primary modeling systematic and will permit an \mbh\ determination with percent-level precision. Based on the Cycle 6 observations presented in this paper, we expect CO imaging with a similar line sensitivity of 0.2 mJy beam\per\ per 20 \kms\ channel and a resolution of 0\farcs1 will allow for a direct determination of the extended mass distribution and a resulting high-precision measurement of the BH mass.

The ALMA observations of NGC 4261 highly resolve \rg\ both along the major and minor axes, and the \cotwo\ and \cothree\ kinematics clearly show Keplerian features. However, the lower S/N of the CO disk leads to degeneracies between \mbh, $i$, and \sigmaturb. Moreover, there are indications that the CO disk is mildly warped with a kinematic PA that varies by $\sim$$15\degr$ over the disk. The NGC 4261 CO kinematics are less affected by dust-correction errors than those for NGC 315. Since the CO emission in NGC 4261 probes deeper within \rg, we expect follow-up observations at $\sim$0.17 mJy beam\per\ per 20 \kms\ channel and $\theta_\mathrm{FWHM}\lesssim0\farcs15$ would facilitate more robust dynamical models that account for the disk's warped structure and intrinsic line widths, and would yield a very precise \mbh\ determination.

\subsection{BH Mass Measurement Prospects for Radio-loud Galaxies with ALMA}

Including the two galaxies presented here, ALMA observations have resolved CO emission in at least 11 FR I ETGs \citep{boizelle17,nagai19,rose19,ruffa19b,ruffa19a,vila19,north19}. NGC 4261 and two other FR I galaxies show central CO emission deficits, with the $\Sigma^\prime{}_\mathrm{CO}$ of the other two tracing large (kpc-scale) rings. The CO emission in the remaining eight FR I galaxies is centrally concentrated in disks with outer radii of $\lesssim$500 pc. Central CO emission is not detected at circular speeds above 700 \kms\ for any FR I galaxy observed to date, which suggests the CO-emitting gas is absent or faint within about 10$-$30 pc of the BH (for more discussion, see \citealp{davis19}; \citetalias{boizelle19}). However, these radio-loud ETGs are expected to have large \rg\ of $\sim$50$-$250 pc assuming the current BH $-$ host galaxy relations, and over half of the ALMA CO data sets show at least some evidence for Keplerian-like gas rotation.

Nearly all of the 11 radio-loud galaxies with CO emission detected by ALMA host dust disks. In one exception \citep[IC 1459;][]{ruffa19a}, filamentary dust is not accompanied by CO emission, and in another case \citep[NGC 1275;][]{nagai19}, the bright nucleus may obscure circumnuclear dust features. Other studies have highlighted the tendency for ETGs with nuclear radio emission to contain dust disks \citep[e.g.,][]{tran01,nyland16}, further suggesting a link between the formation of dust disks and AGN. For example, previous \hst\ surveys detected dust near the centers of about half of all nearby radio-loud galaxies \citep[e.g.,][]{vandokkum95,martel99,verdoes99}, roughly the same as the fraction in radio-quiet ETGs \citep[e.g.,][]{ebneter85,ebneter88,vandokkum95,tomita00,tran01,lauer05}.

Taken together, the large expected \rg\ and the high prevalence of dust/molecular gas disks make FR I galaxies compelling candidates for precision black hole mass measurements. It is worth noting, however, that for galaxies with the brightest compact mm/sub-mm continua \citep[$S_\mathrm{nuc}\gtrsim1$ Jy; e.g.,][]{nagai19}, the spectral dynamic range resulting from calibration and deconvolution errors can limit the faintest detectable emission-line features. Also, CO absorption is sometimes observed against the nuclear continuum \citep[e.g.,][]{boizelle17,nagai19,rose19,ruffa19a} and could dilute emission features. On the plus side, the presence of bright ($>$50 mJy) nuclear mm/sub-mm emission allows for a better phase solution using continuum self-calibration, and CO line imaging with $\theta_\mathrm{FWHM}\lesssim\rg$ can help to disentangle emission and absorption while unambiguously isolating rapid central emission arising from well within \rg.

\section{Conclusion\label{sec:conclusion}}

We present the first dynamical BH mass measurement for NGC 315 and a much improved \mbh\ determination for NGC 4261 based on ALMA $\sim$0\farcs2$-$0\farcs3 resolution CO imaging. With the NGC 315 \cotwo\ and NGC 4261 \cotwo\ and \cothree\ data, we examined the spatial and kinematic structure of the arcsecond-scale circumnuclear disks. We detected CO emission well within the BH-dominated region of the galaxies and traced Keplerian-like rotation down to just 15$-$30 pc (or $\sim$0.1\rg) from their BHs. Using thin-disk gas-dynamical models, we inferred BH masses of $2.08\times10^9$ \msun\ for NGC 315 and $1.67\times10^9$ \msun\ for NGC 4261, which are generally consistent with the predictions from the BH$-$host galaxy relations. In the case of NGC 4261, we have revised the prior ionized gas-dynamical \mbh\ measurement \citep{ferrarese96} upward by a factor of $\sim$3.

We explored statistical uncertainties and various sources of systematic errors to establish BH mass confidence intervals. In both galaxies, the molecular gas is accompanied by significant dust that obscures the stellar light at the center of even near-IR \hst\ images. For the highly inclined disk in NGC 315, we estimated a central dust extinction of $A_J\lesssim1.50$ mag and constructed luminous mass models that bracket the range of likely stellar surface brightness distributions. Adopting the various luminous mass models when fitting gas-dynamical models to the ALMA data cube resulted in a change of $\sim$15\% in \mbh. Dust has a smaller impact on our estimate of \mbh\ for NGC 4261, and we found a negligible change in \mbh\ when using a luminous mass model with a greater central stellar luminosity density. Instead, the NGC 4261 \mbh\ is affected at the $\sim$10$-$25\% level by degeneracies with the inclination angle, the turbulent gas velocity dispersion, and the systemic velocity, which arise due to a mildly warped CO disk observed at moderately low S/N. We also found a $\sim$10\% change in \mbh\ when modeling \cothree\ ALMA data across half the disk. Ultimately, we determined a statistical uncertainty of $0.5$\% and a systematic uncertainty of $7-15$\% in \mbh\ for NGC 315, and a $6$\% statistical uncertainty and a $14-23$\% systematic uncertainty in \mbh\ for NGC 4261. Although not considered in this paper, we note that uncertainties in the galaxy distances introduce an additional systematic to each BH mass error budget that is commensurate with the respective distance uncertainty.

Our work adds to the rapidly growing number of molecular gas-dynamical \mbh\ determinations from ALMA. With this new method of \mbh\ measurement, we are able to consistently obtain precise masses, at the 10$-$20\% level or below, in nearby ETGs, including those that reside at the currently sparsely populated upper end of the $\mbh-\sigma_\star$, $\mbh-L_\mathrm{bul}$, and $\mbh-M_\mathrm{bul}$ correlations. The observed NGC 315 and NGC 4261 CO kinematics warrant even higher angular resolution (and, for NGC 4261, higher S/N) imaging  with ALMA to fully map \vlos\ within \rg, which should enable exceptionally robust BH mass measurements free from the dominant systematics we explore here. Taking an accurate census of BHs across a wide variety of galaxies is essential to understanding the role of BHs in galaxy evolution, and percent-level \mbh\ precision will enable a more thorough exploration of accretion processes for these active galaxies.

\acknowledgments

Based on observations with the NASA/ESA Hubble Space Telescope obtained at the Space Telescope Science Institute, which is operated by the Association of Universities for Research in Astronomy, Incorporated, under NASA contract NAS5-26555. Support for Program number 15909 was provided through a grant from the STScI under NASA contract NAS5-26555. J.~L.~W. was supported in part by NSF grant AST-1814799. Research at UC Irvine was supported by NSF grant AST-1614212. L.~C.~H. was supported by the National Science Foundation of China (11721303, 11991052) and the National Key R\&D Program of China (2016YFA0400702). This paper makes use of the following ALMA data: ADS/JAO.ALMA\#2017.1.00301.S and ADS/JAO.ALMA\#2017.1.01638.S. ALMA is a partnership of ESO (representing its member states), NSF (USA) and NINS (Japan), together with NRC (Canada), MOST and ASIAA (Taiwan), and KASI (Republic of Korea), in cooperation with the Republic of Chile. The Joint ALMA Observatory is operated by ESO, AUI/NRAO and NAOJ. The National Radio Astronomy Observatory is a facility of the National Science Foundation operated under cooperative agreement by Associated Universities, Inc. The work has made use of the NASA/IPAC InfraRed Science Archive, which is funded by NASA and operated by the California Institute of Technology.

\software{AstroDrizzle \citep{gonzaga12}, TweakReg \citep{gonzaga12}, GALFIT \citep{peng02,peng10}, IRAF \citep{tody86,tody93}, CASA \citep[v5.1.2;][]{mcmullin07}}

\appendix

In Table~\ref{tbl:tbla} we list the best-fit parameters of the ``dust-corrected'' MGEs for NGC 315 and NGC 4261. Details of the dust correction and MGE construction are provided in Section~\ref{sec:extinction}.

\begin{deluxetable*}{c|ccc|ccc|ccc}[ht]
\tabletypesize{\small}
\tablecaption{Dust-corrected MGE Parameters\label{tbl:tbla}}
\tablewidth{0pt}
\tablehead{
\multicolumn{1}{c|}{$j$} & 
\colhead{$\log_{10}$ $I_{J,j}$ (\lsun\ pc\pertwo)} & 
\colhead{$\sigma_{j}^\prime{}$ (arcsec)} & 
\multicolumn{1}{c|}{$q_{j}^\prime{}$} & \colhead{$\log_{10}$ $I_{J,j}$ (\lsun\ pc\pertwo)} & 
\colhead{$\sigma_{j}^\prime{}$ (arcsec)} & 
\multicolumn{1}{c|}{$q_{j}^\prime{}$} & \colhead{$\log_{10}$ $I_{H,j}$ (\lsun\ pc\pertwo)} & 
\colhead{$\sigma_{j}^\prime{}$ (arcsec)} & 
\colhead{$q_{j}^\prime{}$}\\[-1.5ex]
\multicolumn{1}{c|}{(1)} & \colhead{(2)} & \colhead{(3)} & \multicolumn{1}{c|}{(4)} & \colhead{(5} & \colhead{(6)} & \multicolumn{1}{c|}{(7)} & \colhead{(8)} & \colhead{(9)} & \colhead{(10)}}
\startdata
  & \multicolumn{6}{c|}{\bf NGC 315} & \multicolumn{3}{c}{\bf NGC 4261}\\
  & \multicolumn{3}{c}{$A_J=0.75$ mag} & \multicolumn{3}{c|}{$A_J=1.50$ mag} & \multicolumn{3}{c}{$A_H=0.40$ mag}\\ \cline{2-4} \cline{5-7} \cline{8-10}
1 & 3.924 & 0.178 & 0.764 & 4.407 & 0.119 & 0.787 & 5.219 & 0.017 & 0.650 \\
2 & 3.896 & 0.617 & 0.716 & 3.912 & 0.644 & 0.681 & 1.753 & 0.102 & 0.650 \\
3 & 3.899 & 1.292 & 0.777 & 3.891 & 1.294 & 0.781 & 4.357 & 1.257 & 0.787 \\
4 & 3.459 & 2.414 & 0.706 & 3.457 & 2.409 & 0.705 & 4.152 & 2.801 & 0.724 \\
5 & 3.474 & 4.159 & 0.722 & 3.476 & 4.152 & 0.722 & 3.667 & 5.222 & 0.710 \\
6 & 3.014 & 8.211 & 0.663 & 3.015 & 8.206 & 0.663 & 3.369 & 9.995 & 0.765 \\
7 & 2.844 & 13.26 & 0.748 & 2.844 & 13.26 & 0.748 & 3.034 & 16.74 & 0.844 \\
8 & 2.072 & 26.50 & 0.765 & 2.073 & 26.49 & 0.763 & 2.409 & 37.29 & 0.817 \\
9 & 2.164 & 30.83 & 0.689 & 2.164 & 30.84 & 0.690 & 2.059 & 65.12 & 0.901 \\
10 & \phantom{*}1.839* & \phantom{*}61.95* & \phantom{*}0.810* & \phantom{*}1.839* & \phantom{*}61.95* & \phantom{*}0.810* & \phantom{*}0.914* & \phantom{*}144.4* & \phantom{*}0.820* \\
11 & \phantom{*}0.939* & \phantom{*}192.6* & \phantom{*}0.980* & \phantom{*}0.939* & \phantom{*}192.6* & \phantom{*}0.980* & \nodata & \nodata & \nodata
\enddata
\begin{singlespace}
  \tablecomments{NGC 315 and NGC 4261 MGE solutions constructed from their respective \hst\ $J$ and $H$-band mosaics. For NGC 315, the model has a uniform $\mathrm{PA}=44.31\degr$ for all components, while the best-fit NGC 4261 MGE has $\mathrm{PA}=-22.03\degr$. Column (1) lists the component number, column (2) is the central surface brightness assuming absolute solar magnitudes of $M_{\odot,\,J}=3.82$ mag and $M_{\odot,\,H}=3.37$ mag (Willmer 2018), column (3) gives the Gaussian standard deviation along the major axis, and column (4) provides the component axis ratio. Primes indicate projected quantities. \spitzer\ IRAC1 MGE components, identified with an asterisk, were included as fixed components when an MGE was fit to the \hst\ image. When constructing the NGC 315 MGE, we modeled the AGN as a point source (with $m_J=19.2$ mag); likewise, the NGC 4261 MGE was accompanied by an unresolved nuclear source (with $m_H=19.6$ mag).}
\end{singlespace}
\end{deluxetable*}

\bibliography{ms.bib}

\end{document}